\documentclass[aip,jcp,preprint,showpacs]{revtex4-1}
\usepackage{graphicx}
\usepackage{multirow}
\usepackage{subfig}

\begin{document}


\title{Carbon dioxide adsorption and activation on Ceria (110): A density functional theory study} 



\author{Zhuo Cheng}
\author{Brent J. Sherman}
\author{Cynthia S. Lo}
\email[]{clo@wustl.edu}
\homepage[]{http://caml.engineering.wustl.edu/}
\affiliation{Department of Energy, Environmental and Chemical Engineering, Washington University, St. Louis, Missouri 63130, USA}


\date{\today}

\begin{abstract}
Ceria (CeO$_2$) is a promising catalyst for the reduction of carbon dioxide (CO$_2$) to liquid fuels and commodity chemicals, in part because of its high oxygen storage capacity, yet the fundamentals of CO$_2$ adsorption and initial activation on CeO$_2$ surfaces remain largely unknown. We use density functional theory, corrected for onsite Coulombic interactions (DFT+U), to explore various adsorption sites and configurations for CO$_2$ on stoichiometric and reduced CeO$_2$ (110).  Our model of reduced CeO$_2$ (110) contains oxygen vacancies at the topmost atomic layer and undergoes surface reconstruction upon introduction of these vacancies.  We find that CO$_2$ adsorption on reduced CeO$_2$ (110) is thermodynamically favored over the corresponding adsorption on stoichiometric CeO$_2$ (110).  The most stable adsorption configuration consists of CO$_2$ adsorbed parallel to the reduced CeO$_2$ (110) surface, with the molecule situated near the site of the oxygen vacancy.  Structural changes in the CO$_2$ molecule are also observed upon adsorption, so that the resulting O-C-O angle is 136.9$^\circ$ and the C-O bonds are 1.198 \AA \ and 1.311 \AA \ in length, respectively.  The molecule bends out of plane to form a unidentate carbonate, as opposed to the bidentate carbonate found by other researchers for CO adsorption to stoichiometric CeO$_2$.  We deduce that charge transfer from reduced surface Ce$^{3+}$ ions to the adsorbate to form the carbonate anion is the first step in the activation and reduction process and cleavage of the C-O bond.
\end{abstract}


\maketitle 

\section{Introduction}
\label{sec:intro}

The synthesis of liquid fuels and commodity chemicals by carbon dioxide (CO$_2$) reforming is a promising approach for clean energy production.  For this reason, much academic and industrial effort has been devoted to exploring efficient means of reducing CO$_2$ \cite{havran_indengchemres_2011}.  It is widely believed that the first step in CO$_2$ reduction is the activation of the C=O bond and eventual formation of the CO$_2^-$ radical \cite{centi_cataltoday_2009}.  Since CO$_2$ is thermodynamically stable and the formation of the radical is difficult in the gas phase \cite{compton_jchemphys_1975} (i.e., -1.90 V vs. NHE \cite{benson_chemsocrev_2009}), we seek appropriate catalysts for lowering the activation barrier to (photo)electrochemical CO$_2$ reduction.

There have been several metal oxides proposed for CO$_2$ activation, such as ZrO$_2$ \cite{kohno_physchemchemphys_2000}, TiO$_2$ \cite{markovits_jmolstructheochem_1996,he_jphyschemc_2010,sorescu_jchemphys_2011a,sorescu_jchemphys_2011b}, MgO \cite{teramura_jphyschemb_2004}, and CaO \cite{besson_surfsci_2012}. Among these various catalysts for CO$_2$ activation, ceria (CeO$_2$) is attracting increased interest due to its high oxygen storage capacities.  CeO$_2$ is commonly used for oxidation and reduction reactions, since Ce ably and reversibly converts between Ce$^{4+}$ and Ce$^{3+}$ upon release and storage of oxygen \cite{trovarelli_book_2002}. 

While industrial methods to efficiently utilize carbon monoxide (CO), which has similar chemistry to CO$_2$, have been developed, CO$_2$ has been comparatively underutilized as a reactant. We already know that the interaction between ceria and CO is vitally important in applications ranging from three-way automotive catalysts (TWC) \cite{kim_indengchemprodrd_1982}, the water-gas shift (WGS) reaction \cite{bunluesin_applcatalb_1998, hilaire_applcatala_2004}, and syngas production \cite{pino_applcatala_2002,pino_applcatala_2003}.  However, all of the research to date has yet to produce a comprehensive model of how ceria and CO$_2$ interact at a molecular level \cite{herschend_chemphys_2006, yang_chemphyslett_2004}, although there have been several experimental studies with ceria as a support for metal nanoparticles that have hinted at important factors; researchers have noted the importance of the interaction between the metal nanoparticle and the oxide support for CO$_2$ activation \cite{bernal_jcatal_2001} and the possibility of oxygen cycling between the support and the adsorbed CO$_2$ species \cite{jin_jphyschem_1987,otsuka_jcatal_1998,buenolopez_applcatalagen_2008,staudt_jcatal_2010}.  In fact, CO is formed by the decomposition of CO$_2$ on partially reduced Pt/ceria catalysts, whereas on fully pre-oxidized (or stoichiometric) ceria, little to no reactivity is observed \cite{staudt_jcatal_2010}. However, these experimental studies did not clarify the mechanism of CO$_2$ activation on ceria nor the structure of the adsorbed species on the surface. Thus, it is important to investigate CO$_2$ activation on ceria at a molecular level.

A number of first-principles studies on ceria have explored the stoichiometric bulk and associated surfaces.  Reduced ceria, with oxygen vacancies at the surface, has also been studied extensively \cite{yang_jchemphys_2004,fabris_physrevb_2005a,burbano_jphyscondensmat_2011}.  It has been reported that the ceria (110) surface is more catalytically active than either the (111) or the (100) surfaces, due to the metastable state of (110) surface, so that the ordering of surface stability is (111) > (110) > (100) \cite{skorodumova_physrevb_2004,herschend_surfsci_2005,herschend_chemphys_2006}.  However, the energy of formation for a single surface oxygen vacancy (with release of $\frac{1}{2}$O$_2$) on stoichiometric ceria (111) is +2.65 eV  \cite{gandugliapirovano_physrevlett_2009}, while it is +2.11 eV for the ideal ceria (110) surface \cite{yang_physletta_2009}.  Although the ceria (110) surface is the second most stable surface among the low index ceria surfaces, it is thus chemically more active than ceria (111).  The creation of oxygen vacancies on the ceria (110) surface requires the least amount of energy, which facilitates the formation of a reduced surface \cite{nolan_surfsci_2005a,nolan_surfsci_2005b}. 

In this paper we present a comprehensive study on the adsorption of CO$_2$ on the stoichiometric and reduced ceria (110) surfaces. A variety of possible adsorption configurations on these two surfaces are presented in terms of structures and energetics, along with charge transfer and electronic property analyses.  Our results provide theoretical evidence that the initial adsorption of ceria surfaces induces CO$_2$ activation to ultimately form the anion radical CO$_2^-$, and the adsorption process is sensitive to surface structure and composition. 

\section{Computational Methods}

All calculations are performed within the framework of density functional theory (DFT) \cite{hohenberg_physrev_1964,kohn_physrev_1965}, as implemented in the Vienna Ab Initio Simulation Package (VASP 5.2) \cite{kresse_physrevb_1993,kresse_compmatersci_1996,kresse_physrevb_1996} and the generalized gradient approximation of Perdew, Burke, and Ernzerhof \cite{perdew_physrevlett_1996} to represent the exchange-correlation energy.  The Projector-Augmented Wave (PAW) method \cite{blochl_physrevb_1994b, kresse_physrevb_1999}, with a 400 eV energy cutoff, was used to describe the wavefunctions of the atomic cores.  The tetrahedron method with Bl{\"o}chl corrections ~\cite{blochl_physrevb_1994a} was used to set the partial occupancies for the orbitals.  While several $k$-point mesh sizes (e.g. $4 \times 4 \times 4$ up to $11 \times 11 \times 11$) were considered, we ultimately used the $6 \times 6 \times 6$ $\Gamma$-centered Monkhorst-Pack $k$-point mesh for bulk ceria to give results that were sufficiently converged (within $1 \times 10^{-5} \ \text{eV}$ using the conjugate gradient method).

The bulk ceria unit cell contains 8 oxygen atoms and 4 cerium atoms in a cubic unit cell with the $Fm\bar{3}m$ symmetry space group; the initial lattice parameter is $\left| \mathbf{a} \right| = 5.411 \ \text{\AA}$ \cite{kummerle_jsolidstatechem_1999}.  After geometry optimization, the bulk lattice parameter increased slightly to $\left| \mathbf{a} \right| = 5.413 \ \text{\AA}$; this compares favorably to previous studies that also follow the DFT approach \cite{skorodumova_physrevb_2001}.

The electronic band structure of bulk ceria contains two energy gaps, as determined by XPS spectra: 1. A 3 eV band gap between the O $2p$ valence band and the Ce $4f$ conduction band, and 2. A 6 eV band gap between the valence band and the Ce $5d$ band \cite{koelling_solidstatecommun_1983,hill_jphyschemsolids_1993,mullins_surfsci_1998}.  Since it is difficult to accurately represent the $4f$ states in ceria using conventional DFT methods \cite{fabris_physrevb_2005a,kresse_physrevb_2005,fabris_physrevb_2005b}, we have chosen to employ the Hubbard $U$ parameter within the GGA+U approach \cite{herbst_physrevb_1978,anisimov_physrevb_1991a}.  The Hubbard $U$ term acts as an on-site Coulombic interaction to properly localize the electrons in these states, which is especially important for the calculations on reduced ceria, with its partially occupied $4f$ states.  We have chosen for our ceria system $U_{eff} = 5 \ \text{eV}$, which is consistent with recommended values previously published in the literature \cite{loschen_physrevb_2007}.

Slab models consisting of 10 atomic layers and 15 \AA \ of vacuum between periodic images in the $z$ direction were cleaved along the (110) plane from the optimized bulk CeO$_2$ unit cell; following optimization, we observe that atomic layer relaxation occurred primarily within the three topmost layers.  Thus, we cleaved the slab in half along the $xy$ plane, so that a $p \left( 2 \times 2 \right)$ slab consisting of 5 atomic layers (overall slab thickness is 7.622 \AA), with the bottom two layers constrained in space,  was used for all subsequent calculations.  The other atoms are free to move in all directions within the fixed lattice.  The exposed surface area is large enough to accommodate a small gas-phase adsorbate (e.g., CO$_2$) without spurious interactions from neighboring images.

\begin{figure}[ht]
\centering
\subfloat[Stoichiometric][Stoichiometric]{
\includegraphics[height=3.5in]{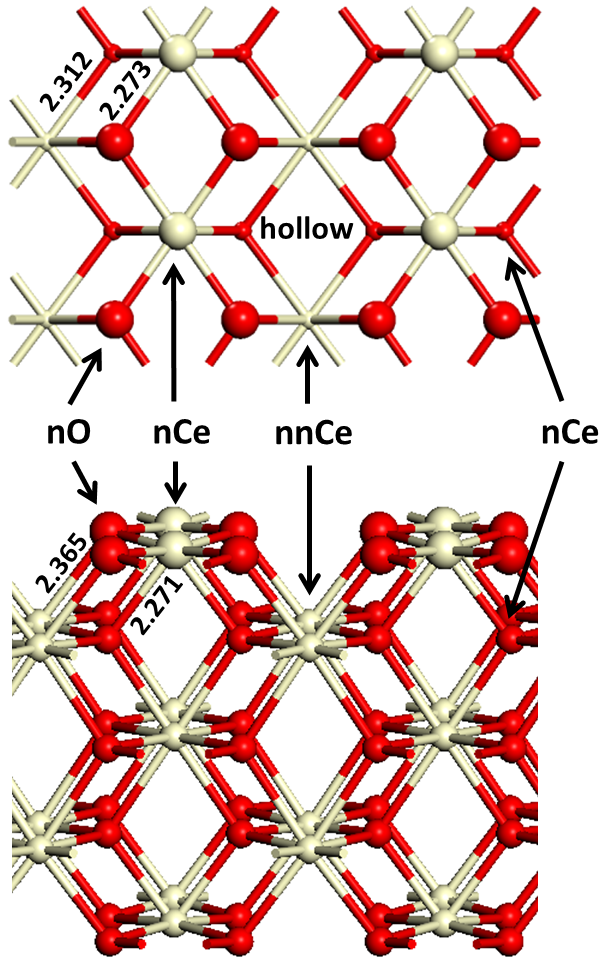}
\label{fig:ceriastoichsurf}
}
\qquad
\subfloat[Reduced][Reduced]{
\includegraphics[height=3.5in]{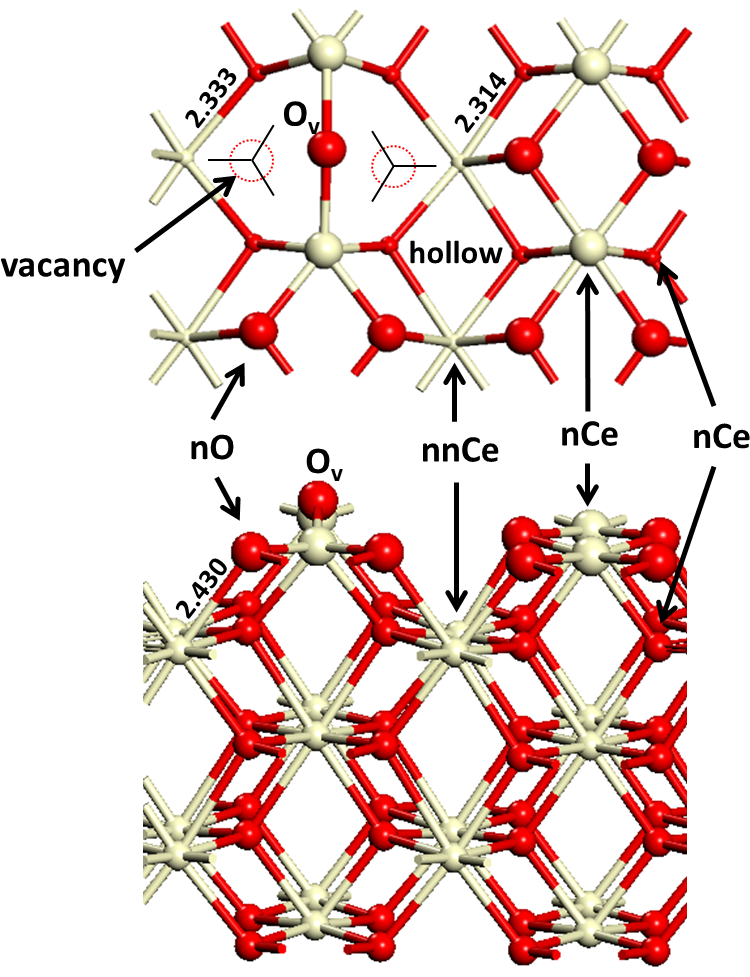}
\label{fig:ceriaredsurf}
}
\caption{Schematic representations of the CeO$_2$ (110) surfaces.  The top figures view the surfaces from the top, and the bottom figures view the surfaces from the side.  nO (large red sphere) refers to a surface O atom.  nCe (large beige sphere) refers to a surface Ce atom.  nnO (small red sphere) refers to a subsurface O atom.  nnCe (small beige sphere) refers to a subsurface Ce atom. }
\end{figure}

The optimized stoichiometric CeO$_2$ (110) surface is shown in Figure \ref{fig:ceriastoichsurf}.  Several distinct adsorption sites are labeled, including: 1. Threefold coordinated O atoms in the topmost atomic layer (nO), 2. Fourfold coordinated O atoms in the subsurface atomic layer (nnO), 3. Sixfold coordinated Ce atoms in the topmost atomic layer (nCe), 4. Eightfold coordinated Ce atoms in the subsurface atomic layer (nnCe), and 5. Hollow.  This nomenclature will be used throughout this discussion to describe changes in the adsorption configuration and electronic structure at the surface.

The optimized reduced CeO$_2$ (110) surface is formed by removing one oxygen atom from the topmost atomic layer of the $p \left( 2 \times 2 \right)$ slab, so that a 12.5\% rate of oxygen vacancies is created; this is shown in Figure \ref{fig:ceriaredsurf}.  Several distinct adsorption sites are labeled, including: 1. Threefold coordinated O atoms in the topmost atomic layer (nO), 2. Fourfold coordinated O atoms in the subsurface atomic layer (nnO), 3. Fivefold and sixfold coordinated Ce atoms in the topmost atomic layer (nCe), 4. Sevenfold and eightfold coordinated Ce atoms in the subsurface atomic layer (nnCe), 5. Hollow, and 6. Vacancy.  We observe that the nO nearest the vacancy site moved to fill the vacancy so that it is equidistant between two nCe atoms; it also moved out of the surface in the perpendicular direction.  Furthermore, the other nnO and nnCe atoms near the vacancy moved slightly away (0.07-0.09 \AA) from the vacancy, so the vacancy site is quite large in area and volume.  As a result, the nCe-O distances near the vacancy site increase to 2.354 \AA, compared to 2.273 \AA \ in stoichiometric CeO$_2$ (110).  This surface reconstruction in reduced CeO$_2$ (110) is in agreement with results previously reported in the literature \cite{campbell_science_2005,fronzi_jchemphys_2009,farmer_science_2010,nolan_chemphyslett_2010}.

The energy of adsorption for carbon dioxide on ceria is given as:
\begin{eqnarray}
E_{ads} & = & E_{\text{CO}_2+\text{CeO}_2 (110)} - E_{\text{CeO}_2 (110)} - E_{\text{CO}_2}
\end{eqnarray}
\noindent where $E_{\text{CeO}_2 (110)}$ is the total energy of the CeO$_2$ (110) surface slab, $E_{\text{CO}_2}$ is the total energy of the CO$_2$ molecule (optimized in a periodic cubic unit cell whose volume is 8000 \AA$^3$), and $E_{\text{CO}_2+\text{CeO}_2 (110)}$ is the total energy of the composite system.  Since the calculations are performed at 0 K and fixed cell volume, the differences in Gibbs free energy should equal the differences in total energy.  By this definition, a negative value of $E_{ads}$ corresponds to an exothermic and spontaneous adsorption process.

Both parallel and perpendicular/vertical orientations of CO$_2$ on the stoichiometric and reduced CeO$_2$ (110) surfaces were considered, and the energy of adsorption calculated for all distinct configurations.  Excess spin density and Bader charge \cite{sanville_jcomputchem_2007} analyses were also performed to study the role of charge transfer in the CO$_2$ adsorption and activation processes at the CeO$_2$ (110) surface.

\section{Results and Discussion}

\subsection{CO$_2$ Adsorption on Stoichiometric CeO$_2$ (110)}

\begin{figure}[htbp]
\centering
\includegraphics[width=6in]{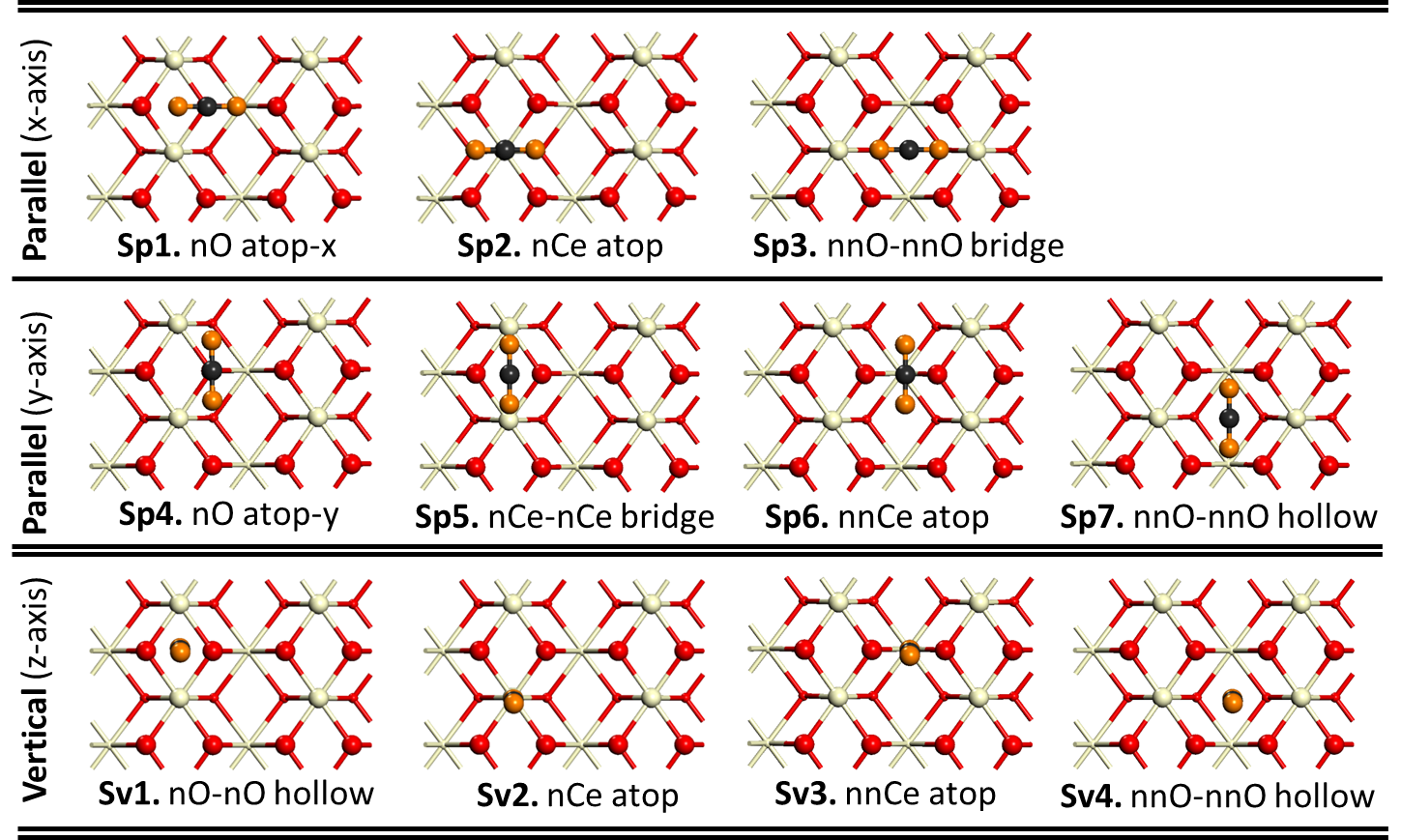}
\caption{CO$_2$ adsorption on $p \left( 2 \times 2 \right)$ supercell of stoichiometric CeO$_2$ (110).  \textbf{Sp} denotes adsorption parallel to the surface, and \textbf{Sv} denotes adsorption perpendicular/vertical to the surface.  For the CO$_2$ adsorbate, oxygen is represented by an orange sphere and carbon is represented by a black sphere.}
\label{fig:co2stoich}
\end{figure}

The optimized adsorption configurations of CO$_2$ on stoichiometric CeO$_2$ (110) are shown in Figure \ref{fig:co2stoich}.  Seven configurations where CO$_2$ is adsorbed parallel to the surface (along either the $x$- or $y$-axis) and four configurations where CO$_2$ is adsorbed perpendicular/vertical to the surface were considered.  The energies of adsorption, select geometrical parameters, and variation of the total Bader charge of the adsorbed CO$_2$ molecule are shown in Table \ref{tab:co2stoich}.  

\begin{table}[htdp]
\begin{center}
\caption{Representative structural parameters, energies of adsorption, and variation of the total Bader charge for a CO$_2$ molecule adsorbed at various sites (Figure \ref{fig:co2stoich}) on the stoichiometric CeO$_2$ (110) surface}
\label{tab:co2stoich}
\begin{tabular}{ccccc}
\hline
\hline
{} & $r_{C-O} \ \left( \text{\AA} \right)$ & $\angle_{O-C-O} \ \left( ^\circ \right)$ & $E_{ads} \ \left( \text{eV} \right)$ & $\frac{\Delta \rho \left( \text{CO}_2 \right)}{\left| e \right|}$ \\
\hline
\textbf{Sp1} & 1.177-1.178 & 180 & -0.034 & -0.003 \\
\textbf{Sp2} & 1.182-1.185 & 178.8 & -0.118 & -0.002 \\
\textbf{Sp3} & 1.183-1.186 & 179.2 & -0.183 & -0.005 \\
\hline
\textbf{Sp4} & 1.178-1.182 & 180 & -0.078 & -0.005 \\
\textbf{Sp5} & 1.196-1.207 & 178.2 & -0.262 & -0.008 \\
\textbf{Sp6} & 1.180-1.182 & 179.5 & -0.043 & -0.002 \\
\textbf{Sp7} & 1.179-1.183 & 178.1 & -0.174 & -0.002 \\
\hline
\textbf{Sv1} & 1.172-1.180 & 178.2 & -0.201 & -0.006 \\
\textbf{Sv2} & 1.173-1.178 & 179.1 & -0.102 & -0.003 \\
\textbf{Sv3} & 1.175-1.176 & 180 & -0.035 & -0.001 \\
\textbf{Sv4} & 1.172-1.179 & 180 & -0.253 & -0.003 \\
\hline
\hline
\end{tabular}
\end{center}
\end{table}

\begin{figure}[htbp]
\centering
\includegraphics[width=6in]{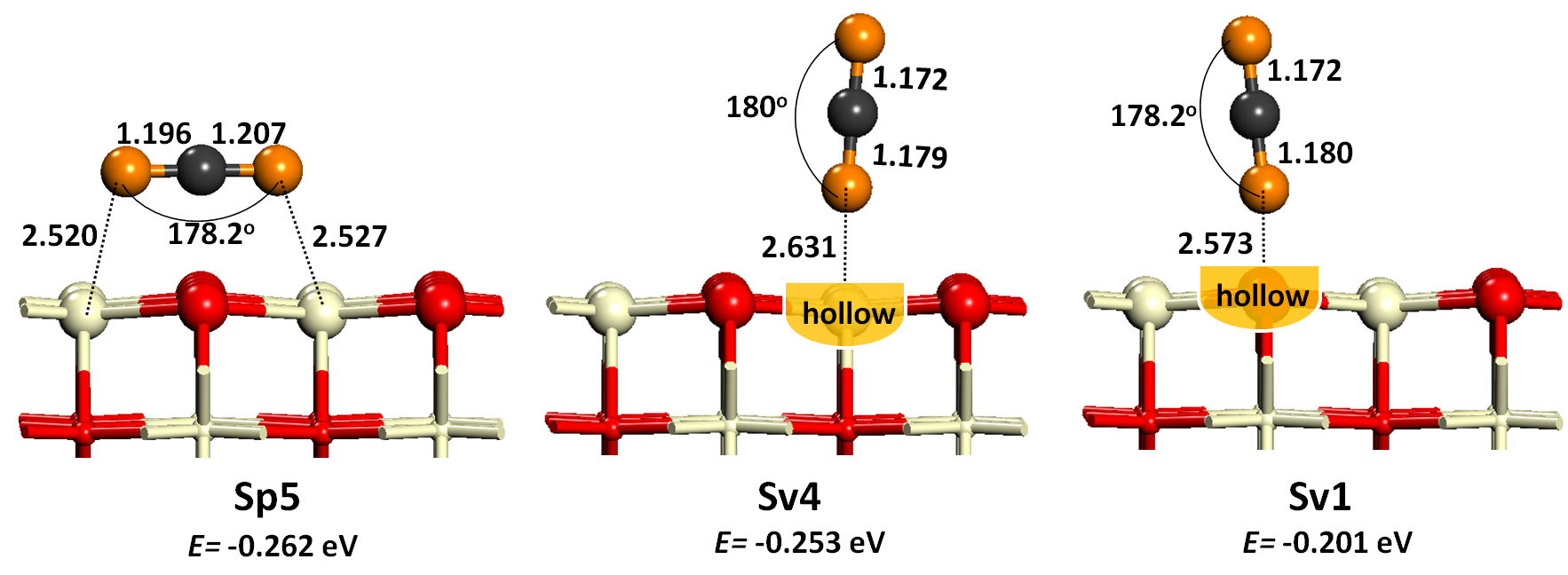}
\caption{CO$_2$ adsorption on stoichiometric CeO$_2$ (110) in selected configurations: \textbf{Sp5}, \textbf{Sv4}, and \textbf{Sv1}.  Both structural parameters and energies of adsorption are depicted.}
\label{fig:co2stoichside}
\end{figure}

The strongest energy of adsorption ($E_{ads} = -0.262 \ \text{eV}$) is observed for nnO adsorbing as a "bridge" between two nCe atoms (\textbf{Sp5}). The two C-O bonds in CO$_2$ are slightly elongated to 1.196 \AA \ and 1.207 \AA, compared to 1.176 \AA \ in the gas phase. The CO$_2$ molecule are nearly linear, with a O-C-O angle of 178.2$^\circ$. The two oxygen atoms in CO$_2$ reside at the top of two sixfold coordinated surface nCe atoms, and the Ce-O distances are 2.520 \AA \ and 2.527 \AA, respectively. An additional depiction of this adsorption configuration (\textbf{Sp5}) is shown in Figure \ref{fig:co2stoichside}. 

The \textbf{Sv4} and \textbf{Sv1} adsorption configurations are slightly less energetically favorable ($E_{ads} = -0.253 \ \text{eV}$ and $E_{ads} = -0.201 \ \text{eV}$, respectively), and both are in vertical orientation and tilted slightly relative to the surface normal. For \textbf{Sv4}, the CO$_2$ adsorbs at the nnO\_nnO hollow site, which is flanked by two subsurface ceria atoms (nnCe) and two subsurface oxygen atoms (nnO). The distance between the ceria surface and the adsorbing oxygen atom in CO$_2$ is 2.631 \AA. For \textbf{Sv1}, the CO$_2$ adsorbs at the nO\_nO hollow site, which is flanked by two surface ceria atoms (nCe) and two surface oxygen atoms (nO). The distance between the ceria surface and the adsorbing oxygen atom in CO$_2$ is 2.573 \AA. The symmetry of these two hollow sites reduces the net intermolecular forces acting on the CO$_2$ molecule, so these adsorption configurations are relatively stable. However, the lengths of the two C-O bonds in CO$_2$ for the \textbf{Sv4} and \textbf{Sv1} configurations are similar to those in gas-phase CO$_2$.  Additional depictions of these these two adsorption configurations are shown in Figure \ref{fig:co2stoichside}. 

Several other adsorption configurations merit discussion, in terms of alignment of the adsorbate and consideration of surface/subsurface effects.  For \textbf{Sp1} and \textbf{Sp4}, the C atom of CO$_2$ resides on top of the surface nO atom, but is aligned along different axes ($x$ and $y$, respectively). Both \textbf{Sp1} and \textbf{Sp4} result in very weak physisorption, with \textbf{Sp4} being slightly favored energetically due to the increased electrostatic repulsion with the neighboring surface nO oxygen in the \textbf{Sp1} configuration compared to the repulsion with the subsurface nnO oxygens in the \textbf{Sp4} configuration.  For \textbf{Sp2} and \textbf{Sv2}, where CO$_2$ is adsorbed on top of the surface nCe atom, the energy of adsorption is stronger than that for \textbf{Sp6} and \textbf{Sv3}, respectively, where CO$_2$ is adsorbed on top of the subsurface nnCe atom.  Any differences in electrostatic repulsion with subsurface nnCe or nnO atoms appear to be minimal, as the energies of adsorption for the \textbf{Sp3} and \textbf{Sp7} configurations are nearly identical.  Thus, we propose that only surface atoms appear to significantly influence the strength of CO$_2$ adsorption to the stoichiometric CeO$_2$ (110) surface.

In general, all of these adsorption configurations on stoichiometric CeO$_2$ (110) would be classified as weak physisorption, since the CO$_2$ molecule remains physically unchanged upon adsorption and the distance between CO$_2$ and the surface is more than 2.52 \AA, which is too far to form a Ce-O bond.  Furthermore, the Bader charge analysis indicates that the variation of the total charge of adsorbed CO$_2$ is less than $-0.01 \left| e \right|$, so there is no significant change in the electronic structure.  We can conclude that stoichiometric CeO$_2$ (110) does not activate the CO$_2$ molecule.

\subsection{CO$_2$ Adsorption on Reduced CeO$_2$ (110)}

\begin{figure}[htbp]
\centering
\includegraphics[width=6in]{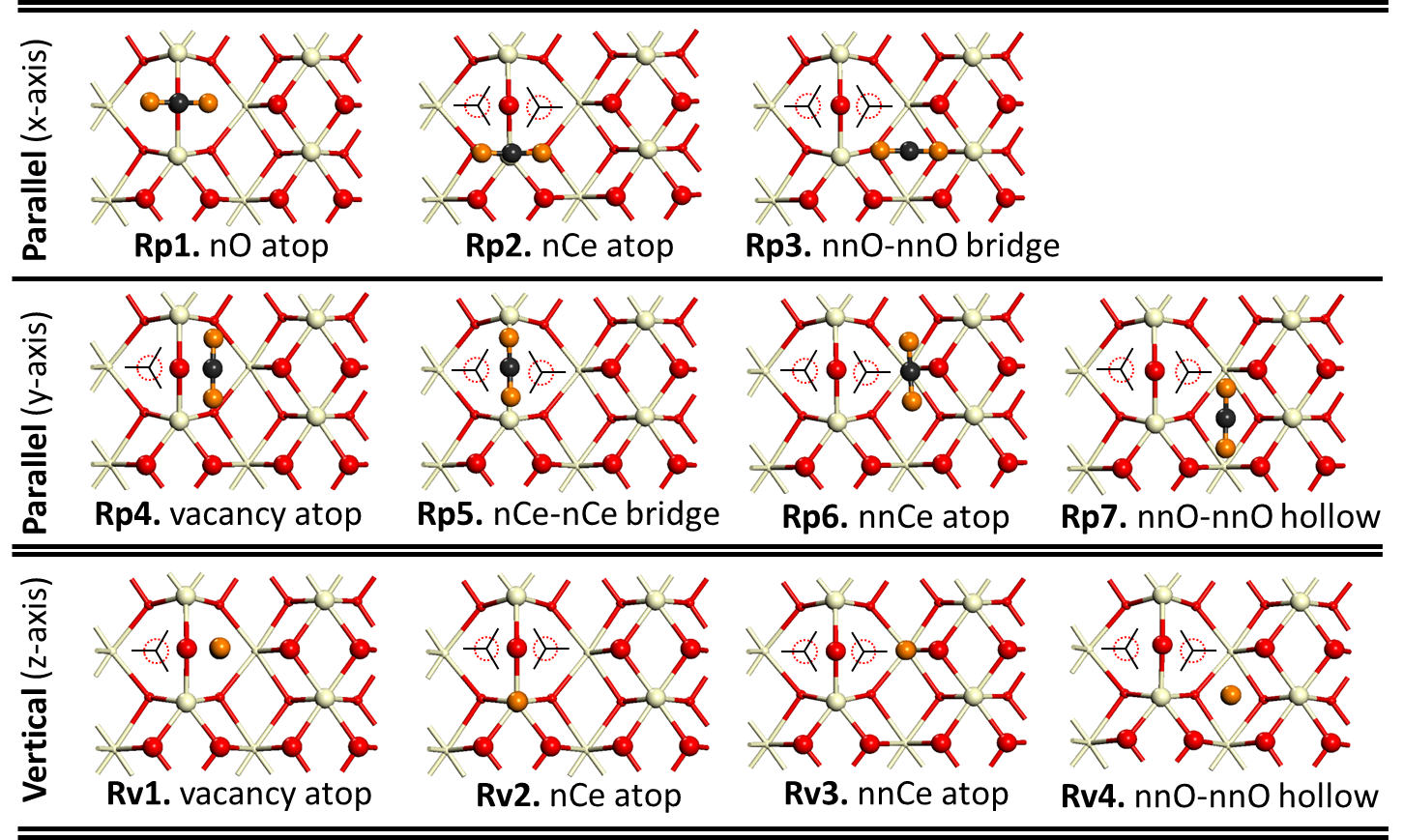}
\caption{CO$_2$ adsorption on $p \left( 2 \times 2 \right)$ supercell of reduced CeO$_2$ (110).  \textbf{Rp} denotes adsorption parallel to the surface, and \textbf{Rv} denotes adsorption perpendicular/vertical to the surface.  For the CO$_2$ adsorbate, oxygen is represented by an orange sphere and carbon is represented by a black sphere.}
\label{fig:co2red}
\end{figure}

The optimized adsorption configurations of CO$_2$ on reduced CeO$_2$ (110) are shown in Figure \ref{fig:co2red}.  Seven configurations where CO$_2$ is adsorbed parallel to the surface (along either the $x$- or $y$-axis) and four configurations where CO$_2$ is adsorbed perpendicular/vertical to the surface were considered.  The energies of adsorption, select geometrical parameters, and variation of the total Bader charge of the adsorbed CO$_2$ molecule are shown in Table \ref{tab:co2red}.  

\begin{table}[htdp]
\begin{center}
\caption{Representative structural parameters, energies of adsorption, and variation of the total Bader charge for a CO$_2$ molecule adsorbed at various sites (Figure \ref{fig:co2red}) on the reduced CeO$_2$ (110) surface}
\label{tab:co2red}
\begin{tabular}{ccccc}
\hline
\hline
{} & $r_{C-O} \ \left( \text{\AA} \right)$ & $\angle_{O-C-O} \ \left( ^\circ \right)$ & $E_{ads} \ \left( \text{eV} \right)$ & $\frac{\Delta \rho \left( \text{CO}_2 \right)}{\left| e \right|}$ \\
\hline
\textbf{Rp1} & 1.176-1.183 & 178.3 & -0.401 & -0.273 \\
\textbf{Rp2} & 1.182-1.197 & 135.5 & -0.119 & -0.223 \\
\textbf{Rp3} & 1.207-1.287 & 177.3 & -0.201 & -0.119 \\
\hline
\textbf{Rp4} & 1.198-1.311 & 140.4 & -1.012 & -0.936 \\
\textbf{Rp5} & 1.246-1.256 & 136.9 & -1.223 & -0.955 \\
\textbf{Rp6} & 1.179-1.183 & 180 & -0.055 & -0.016 \\
\textbf{Rp7} & 1.177-1.182 & 178.9 & -0.177 & -0.059 \\
\hline
\textbf{Rv1} & 1.168-1.182 & 177.1 & -0.815 & -0.081 \\
\textbf{Rv2} & 1.201-1.322 & 139.9 & -0.265 & -0.921 \\
\textbf{Rv3} & 1.172-1.181 & 179.1 & -0.038 & -0.043 \\
\textbf{Rv4} & 1.175-1.182 & 178.5 & -0.282 & -0.178 \\
\hline
\hline
\end{tabular}
\end{center}
\end{table}

\begin{figure}[htbp]
\centering
\includegraphics[width=6in]{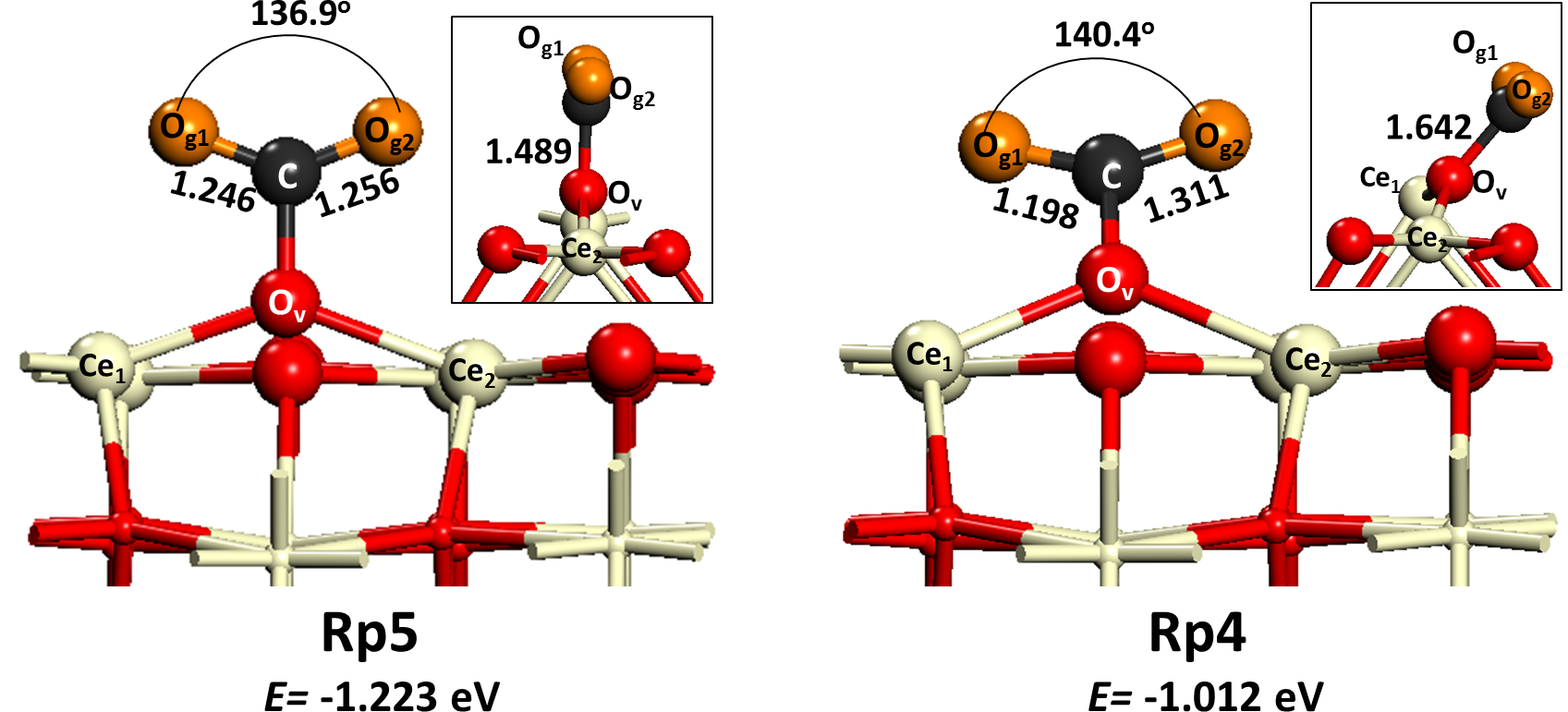}
\caption{CO$_2$ adsorption on reduced CeO$_2$ (110) in selected configurations: \textbf{Rp5} and \textbf{Rp4}. Both structural parameters and energies of adsorption are depicted.}
\label{fig:co2redside}
\end{figure}

The strongest energy of adsorption ($E_{ads} = -1.233 \ \text{eV}$) is observed for CO$_2$ parallel adsorption as a "bridge" between two nCe atoms (\textbf{Rp5}) adjacent to the vacancy site.  From Figure \ref{fig:co2red}, we can see the carbon atom of the adsorbate is situated directly above the nO atom that had moved toward the vacancy oxygen atom.  Furthermore, the adsorbed CO$_2$ molecule exhibits a bent configuration, where the O-C-O angle is 136.9$^\circ$.  Also, the two C-O bonds in the adsorbed CO$_2$ molecule are elongated to 1.246 \AA \ and 1.256 \AA, compared to 1.176 \AA \ in the gas phase.  Thus, we would classify this process as chemisorption.  Although there is significant lengthening of the C-O bonds, complete dissociation to form CO+O is not observed. In addition, the C-O distance between adsorbed CO$_2$ and the surface nO atom is 1.498 \AA, so a carbonate group appears likely but does not actually form upon surface adsorption.  An additional depiction of this adsorption configuration (\textbf{Rp5}) is shown in Figure \ref{fig:co2redside}.

The next strongest energy of adsorption ($E_{ads} = -1.012 \ \text{eV}$) is observed for CO$_2$ parallel adsorption on top of the vacancy (\textbf{Rp4}). The two C-O bonds in the adsorbed CO$_2$ molecule are 1.198 \AA \ and 1.311 \AA, respectively. The adsorbed CO$_2$ molecule exhibits a similar binding geometry and bent configuration with \textbf{Rp5}; however, the O-C-O angle is slightly larger (140.4$^\circ$) than the corresponding O-C-O angle in \textbf{Rp5} (136.9$^\circ$). Furthermore, the C-O distance between adsorbed CO$_2$ and the vacancy nO atom in \textbf{Rp4} is also larger (1.642 \AA), than the corresponding C-O distance in \textbf{Rp5}, which suggests that the adsorbed CO$_2$ molecule is not as activated as in the \textbf{Rp5} configuration.  An additional depiction of this adsorption configuration (\textbf{Rp4}) is shown in Figure \ref{fig:co2redside}.

Vertical adsorption atop the vacancy site (\textbf{Rv1}) is slightly less energetically favorable ($E_{ads} = -0.815 \ \text{eV}$) than for \textbf{Rp5} or \textbf{Rp4}, and we observe no analogous activation of the CO$_2$ molecule; the resulting O-C-O angle is 177.1$^\circ$ and no significant changes are measured in the C-O bond lengths compared to the gas phase values.  Though one oxygen atom of the CO$_2$ molecule is inserted into the surface vacancy, the reduced surface is sufficiently reconstructed so that the remaining surface oxygen has moved to bridge the adjacent surface cerium atoms; thus, the CO$_2$ molecule in the \textbf{Rv1} configuration cannot fully "heal" the vacancy to regenerate the stoichiometric surface.

Vertical adsorption atop the nCe atom nearest the vacancy site (\textbf{Rv2}) is less energetically favorable ($E_{ads} = -0.265 \ \text{eV}$) than adsorption atop the vacancy site (\textbf{Rv1}), although the adsorbed molecule is bent similarly to \textbf{Rp5}.  The other adsorption configurations on reduced CeO$_2$ (110), where CO$_2$ is adsorbed far from the vacancy site, are less energetically stable, with energies of adsorption on the same order of magnitude as those on stoichiometric CeO$_2$ (110); thus, we will not explore these configurations further, and instead, primarily focus our subsequent discussion and analysis on the most stable adsorption configuration, \textbf{Rp5}.  

\subsection{Electronic analysis of CeO$_2$ as a catalyst for CO$_2$ activation}

We now comment on the structural changes induced by oxygen vacancy formation at reduced CeO$_2$ (110) surfaces.  We see that for the \textbf{Rp5} configuration, the Ce-O distances between either neighboring nCe atom and the nO atom at the vacancy are significantly longer (2.292-2.237 \AA) than the corresponding Ce-O distance in the clean reduced surface (2.165 \AA).  This suggests that structural changes occur even following CO$_2$ adsorption, and may be attributable to charge transfer between the surface and adsorbed CO$_2$.  

\begin{figure}[htbp]
\centering
\subfloat[Clean][Clean]{
\includegraphics[width=3in]{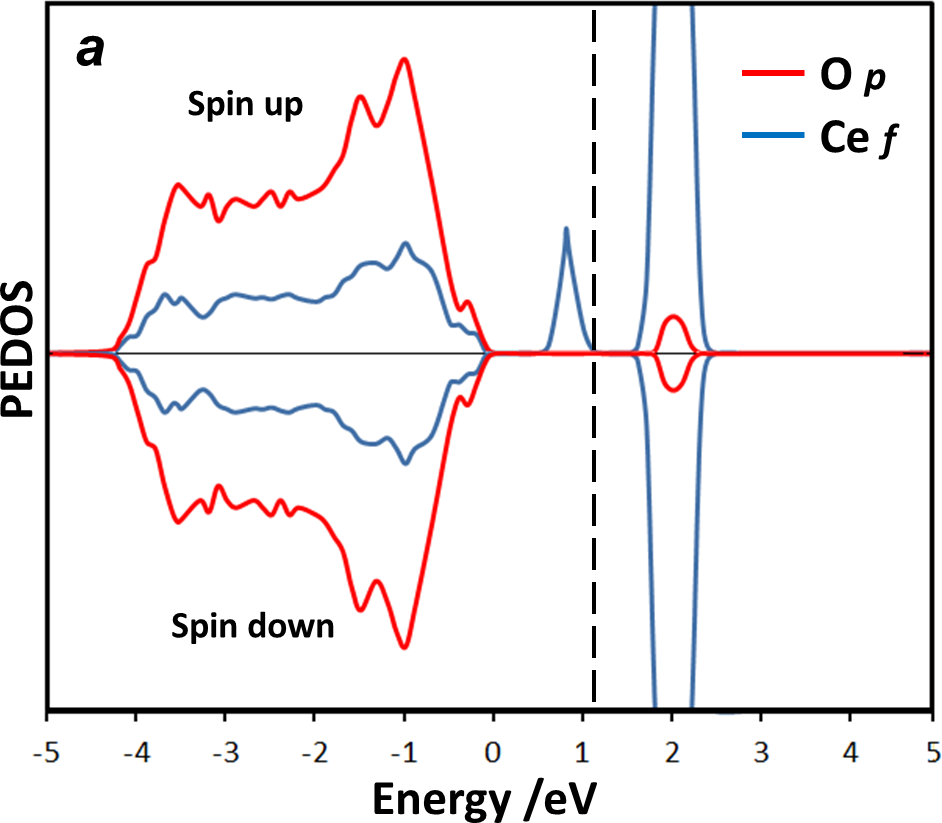}
\label{fig:pedosclean}
}
\qquad
\subfloat[With adsorbed CO$_2$][With adsorbed CO$_2$]{
\includegraphics[width=3in]{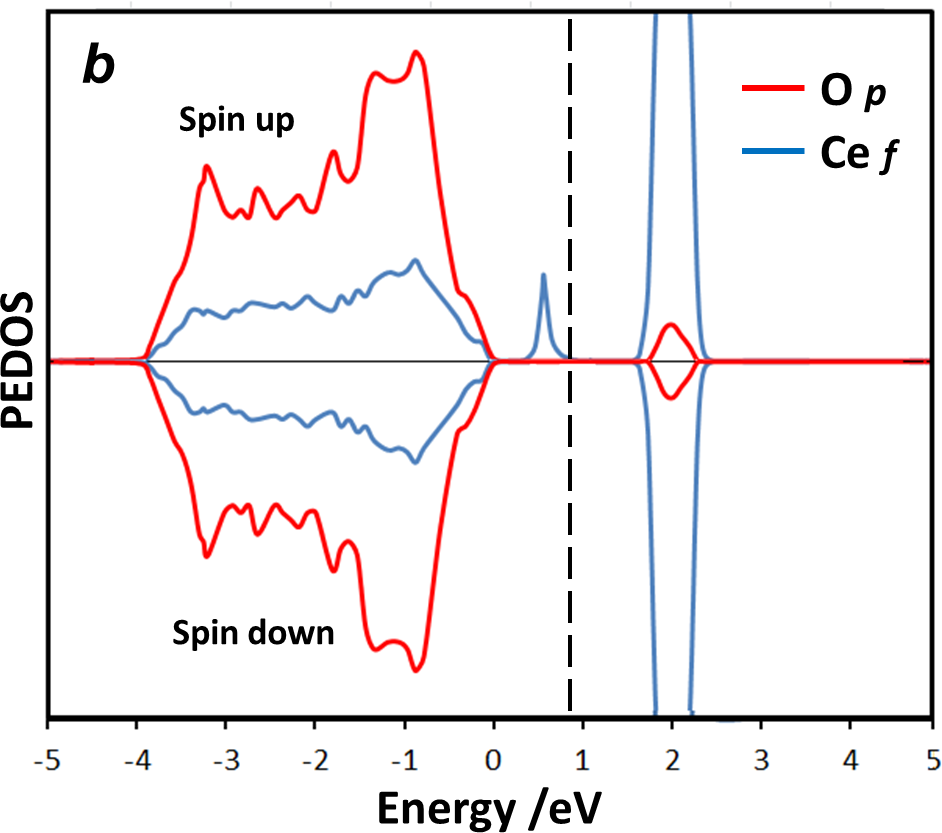}
\label{fig:pedosco2}
}
\caption{Partial electronic density of states (PEDOS) for Ce and O in reduced CeO$_2$ (110). The top of the valence band is aligned to 0 eV, and the Fermi level is indicated with a vertical dashed line.}
\end{figure}

\begin{figure}[ht]
\centering
\subfloat[Clean][Clean]{
\includegraphics[width=3in]{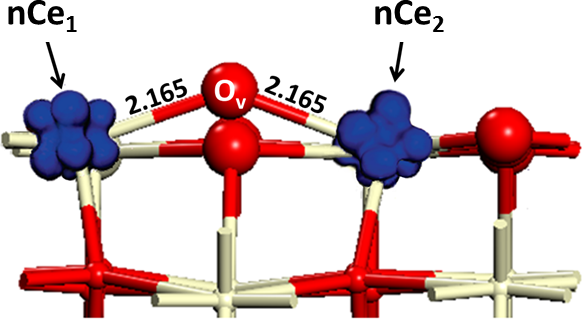}
\label{fig:spinclean}
}
\qquad
\subfloat[With adsorbed CO$_2$][With adsorbed CO$_2$]{
\includegraphics[width=3in]{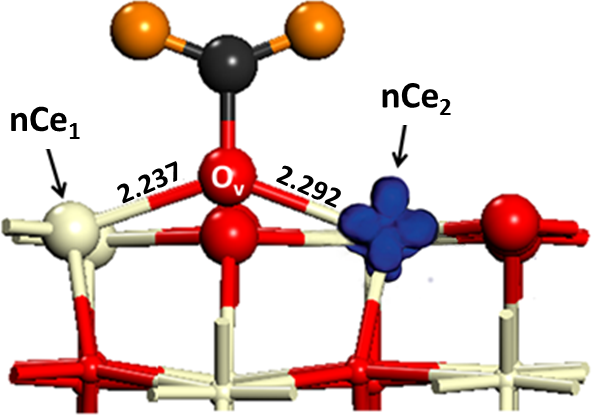}
\label{fig:spinco2}
}
\caption{Excess spin density in reduced CeO$_2$ (110).  Upon formation of the oxygen vacancy, adjacent cerium atoms are reduced to Ce$^{3+}$.  Upon CO$_2$ adsorption, one cerium cation is re-oxidized back to Ce$^{4+}$ and the excess charge is transferred to the adsorbate.  The Ce-O bonds at the surface are also lengthened.}
\label{fig:spin}
\end{figure}

The partial electronic density of states (PEDOS) for Ce and O, when we have the clean reduced CeO$_2$ (110) surface, is shown in Figure \ref{fig:pedosclean}, and it is consistent with those previously published in the literature \cite{nolan_surfsci_2005a,nolan_surfsci_2005b,galea_molsimulat_2009}.  We aligned the top of the valence band to 0 eV, and observed a filled gap state in the middle of the band gap; this filled gap corresponds to the unoccupied Ce $4f$ state. The integration of this peak indicates that there are two electrons per oxygen vacancy. To locate these two electrons, we plotted the excess spin density, which is defined as the difference between the up spin density and the down spin density, for clean reduced CeO$_2$ (110), as shown in Figure \ref{fig:spinclean}.  We see that the spin density on the clean reduced CeO$_2$ (110) surface is localized on the two nCe atoms on the surface nearest the vacancy. The shapes of the two isosurfaces resemble those of $f$ orbitals, which is consistent with the $4f$ electronic configuration in Ce$^{3+}$. Since cerium exists in the Ce$^{4+}$ state in stoichiometric CeO$_2$, the excess spin density plot suggests that it adopts the Ce$^{3+}$ state upon formation of the oxygen vacancy.  This result is also in agreement with previous observations on other reduced ceria surfaces \cite{henderson_surfsci_2003}.

The PEDOS for Ce and O in the situation where CO$_2$ is adsorbed to reduced CeO$_2$ (110) is shown in Figure \ref{fig:pedosco2}.  We observe that the gap between the Fermi level (top of the unoccupied Ce $4f$ peak) and the bottom of the conduction band widens upon CO$_2$ adsorption, and nears the value observed for stoichiometric CeO$_2$ (110) \cite{yang_jchemphys_2004}.  In addition, the filled gap state peak narrows upon CO$_2$ adsorption; the integration of this peak indicates that there is only one electron per oxygen vacancy, compared to two electrons in the clean reduced surface.  To locate the remaining electron, we plotted the excess spin density for reduced CeO$_2$ (110) with adsorbed CO$_2$ (\textbf{Rp5}), as shown in Figure \ref{fig:spinco2}. From Figure \ref{fig:spinco2}, we observe that the remaining electron is localized on the nCe$_2$ atom; thus one reduced Ce$^{3+}$ cation (nCe$_2$) remains on the surface, while the other (nCe$_1$) has been re-oxidized back to Ce$^{4+}$. The asymmetric electric field at the surface vacancy, with one Ce$^{3+}$ and one Ce$^{4+}$, explains the asymmetric C-O bond elongation found in the most stable adsorption configuration (\textbf{Rp5}). We also observe that the re-oxidized Ce$^{4+}$ atom is closer in distance to the nO oxygen atom that moved towards the vacancy site upon surface reduction, while the Ce$^{3+}$ atom is further from this nO oxygen atom and the adsorbed CO$_2$ molecule due to electrostatic repulsion.  These results indicate that the reduced ceria (110) surface has been partially oxidized upon CO$_2$ adsorption at the vacancy site. 

\begin{figure}[htbp]
\centering
\includegraphics[width=3in]{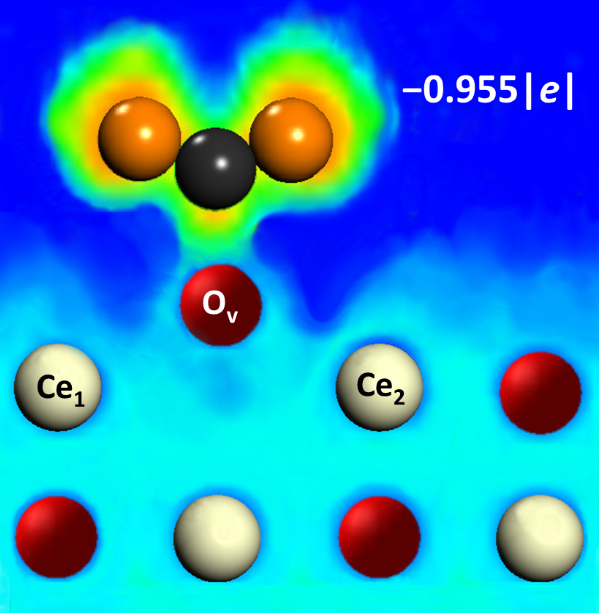}
\caption{Electron localization function for reduced CeO$_2$ (110) with adsorbed CO$_2$ (\textbf{Rp5}).  A net charge of $-0.955 \left| e \right|$ is transferred from the surface to form an adsorbed unidentate carbonate species.}
\label{fig:elf}
\end{figure}

Since the CO$_2$ molecule may accept electrons into its lowest unoccupied molecular orbital to form the carbonate anion, we performed a Bader charge analysis on the system, focusing on the interface between CO$_2$ and the reduced CeO$_2$ (110) surface, and plotted the electron localization function (ELF) \cite{becke_jchemphys_1990}.  From the variation of the Bader charges on the CO$_2$ molecule in various adsorption configurations (Table \ref{tab:co2stoich}-Table \ref{tab:co2red}), we observe definitive charge transfer from the ceria surface to the adsorbed CO$_2$ molecule.  This is especially prominent in the case of the \textbf{Rp5} configuration shown in Figure \ref{fig:spin}, where a net charge of $-0.955 \left| e \right|$ is transferred to form an adsorbed unidentate carbonate species.  This is confirmed by the ELF plot shown in Figure \ref{fig:elf}, where it is clear that the adsorbed CO$_2$ is activated with a net negative charge localized on the oxygen atoms of the molecule.  Thus, CO$_2$ is activated to form a unidentate carbonate anion upon adsorption parallel to vacancy sites on reduced CeO$_2$ (110) surfaces.

We also performed a Bader charge analysis for the perpendicular/vertical adsorption configurations of CO$_2$ to the reduced CeO$_2$ (110) surface.  We observed that while no significant charge transfer occurs for the \textbf{Rv1} configuration, the variation in the Bader charge for the \textbf{Rv2} configuration is $-0.921 \left| e \right|$, which again suggests that Ce$^{3+}$ is oxidized back to Ce$^{4+}$ when a CO$_2$ molecule available for reduction is placed in close proximity.  However, the low energy of adsorption for the \textbf{Rv2} configuration suggests that it is not favored compared to the \textbf{Rp5} configuration.

In summary, we conclude that the specific structure of the ceria surface has an critical influence on the CO$_2$ activation. CO$_2$ may chemisorb to reduced CeO$_2$ (110) with an energy of adsorption of -1.223 eV, and that charge transfer from the reduced ceria surface to the adsorbate results in the formation of an activated unidentate carbonate species.  In the CO$_2$ adsorption process,  the reduced ceria (110) surface is only partially re-oxidized, which suggests that there is additional capacity for further reaction if the CO$_2$ concentration at the surface is increased. Our present findings are in agreement with recent experimental work, in which the researchers showed that re-oxidation of ceria surfaces using CO$_2$ would only be energetically feasible at the adsorption sites that have been sufficiently reduced \cite{staudt_jcatal_2010}.  Furthermore, partial re-oxidation of reduced ceria occurs with high reaction probability even at room temperature \cite{staudt_jcatal_2010}, so we propose that reduced ceria is a promising catalyst for CO$_2$ activation and further reaction.

\section{Conclusions}

We have performed analyses of structural geometries, energetics and electronic properties for the process of carbon dioxide adsorption to stoichiometric and reduced ceria surfaces, using the framework of density functional theory and employing the Hubbard $U$ correction.  It is found that oxygen vacancy plays an important role in the adsorption of CO$_2$ on the ceria surface.  While CO$_2$ adsorption to stoichiometric CeO$_2$ (110) is relatively weak, with no significant changes to the geometry or electronic structure of the adsorbate, CO$_2$ adsorption to reduced CeO$_2$ (110) is favored, particularly in the case of parallel CO$_2$ adsorption atop the vacancy oxygen site neighboring two reduced Ce$^{3+}$; the energy of adsorption is -1.223 eV.  This chemisorption results in a bent CO$_2$ molecule with an O-C-O angle of 136.9$^\circ$ and elongated C-O bonds compared to the gas-phase molecule.  Bader charge analysis confirms that a net charge of $-0.955 \left| e \right|$ is transferred from the reduced CeO$_2$ (110) surface to the adsorbed CO$_2$ molecule in the most stable adsorption structure. Thus, the adsorbed CO$_2$ is activated to form the unidentate carbonate anion with the net negative charge localized on the oxygen atoms of the molecule.  The excess spin density shows that the reduced ceria surface is partially re-oxidized. The re-oxidized Ce$^{4+}$ atom is closer in distance to the nO oxygen atom that moved towards the vacancy site upon surface reduction, while the Ce$^{3+}$ atom is further from this nO oxygen atom and the adsorbed carbonate anion due to electrostatic repulsion. We suggest that reduced CeO$_2$ (110) is a promising catalyst for either thermal or electrochemical reduction of CO$_2$ as the first step in the formation of fuels and chemicals for the closing of the carbon cycle.


%
%
%

\begin{acknowledgments}
This material is based upon work supported by the Consortium for Clean Coal Utilization at Washington University in St. Louis, and by the National Science Foundation through Teragrid resources provided by the Texas Advanced Computing Cluster under grant number TG-CTS100011.  The authors acknowledge Milorad Dudukovic, John Gleaves, Vesna Havran, Gregory Yablonsky, and Sandra Matteucci for insightful comments and discussions.
\end{acknowledgments}

\bibliography{cheng_arxiv_2012}

\begin{thebibliography}{60}%
\makeatletter
\providecommand \@ifxundefined [1]{%
 \@ifx{#1\undefined}
}%
\providecommand \@ifnum [1]{%
 \ifnum #1\expandafter \@firstoftwo
 \else \expandafter \@secondoftwo
 \fi
}%
\providecommand \@ifx [1]{%
 \ifx #1\expandafter \@firstoftwo
 \else \expandafter \@secondoftwo
 \fi
}%
\providecommand \natexlab [1]{#1}%
\providecommand \enquote  [1]{``#1''}%
\providecommand \bibnamefont  [1]{#1}%
\providecommand \bibfnamefont [1]{#1}%
\providecommand \citenamefont [1]{#1}%
\providecommand \href@noop [0]{\@secondoftwo}%
\providecommand \href [0]{\begingroup \@sanitize@url \@href}%
\providecommand \@href[1]{\@@startlink{#1}\@@href}%
\providecommand \@@href[1]{\endgroup#1\@@endlink}%
\providecommand \@sanitize@url [0]{\catcode `\\12\catcode `\$12\catcode
  `\&12\catcode `\#12\catcode `\^12\catcode `\_12\catcode `\%12\relax}%
\providecommand \@@startlink[1]{}%
\providecommand \@@endlink[0]{}%
\providecommand \url  [0]{\begingroup\@sanitize@url \@url }%
\providecommand \@url [1]{\endgroup\@href {#1}{\urlprefix }}%
\providecommand \urlprefix  [0]{URL }%
\providecommand \Eprint [0]{\href }%
\providecommand \doibase [0]{http://dx.doi.org/}%
\providecommand \selectlanguage [0]{\@gobble}%
\providecommand \bibinfo  [0]{\@secondoftwo}%
\providecommand \bibfield  [0]{\@secondoftwo}%
\providecommand \translation [1]{[#1]}%
\providecommand \BibitemOpen [0]{}%
\providecommand \bibitemStop [0]{}%
\providecommand \bibitemNoStop [0]{.\EOS\space}%
\providecommand \EOS [0]{\spacefactor3000\relax}%
\providecommand \BibitemShut  [1]{\csname bibitem#1\endcsname}%
\let\auto@bib@innerbib\@empty
\bibitem [{\citenamefont {Havran}, \citenamefont {Dudukovi{\'c}},\ and\
  \citenamefont {Lo}(2011)}]{havran_indengchemres_2011}%
  \BibitemOpen
  \bibfield  {author} {\bibinfo {author} {\bibfnamefont {V.}~\bibnamefont
  {Havran}}, \bibinfo {author} {\bibfnamefont {M.~P.}\ \bibnamefont
  {Dudukovi{\'c}}}, \ and\ \bibinfo {author} {\bibfnamefont {C.~S.}\
  \bibnamefont {Lo}},\ }\href {\doibase 10.1021/ie2000192} {\bibfield
  {journal} {\bibinfo  {journal} {Industrial \& Engineering Chemistry
  Research}\ }\textbf {\bibinfo {volume} {50}},\ \bibinfo {pages} {7089}
  (\bibinfo {year} {2011})},\ \Eprint
  {http://arxiv.org/abs/http://pubs.acs.org/doi/pdf/10.1021/ie2000192}
  {http://pubs.acs.org/doi/pdf/10.1021/ie2000192} \BibitemShut {NoStop}%
\bibitem [{\citenamefont {Centi}\ and\ \citenamefont
  {Perathoner}(2009)}]{centi_cataltoday_2009}%
  \BibitemOpen
  \bibfield  {author} {\bibinfo {author} {\bibfnamefont {G.}~\bibnamefont
  {Centi}}\ and\ \bibinfo {author} {\bibfnamefont {S.}~\bibnamefont
  {Perathoner}},\ }\href {\doibase 10.1016/j.cattod.2009.07.075} {\bibfield
  {journal} {\bibinfo  {journal} {Catalysis Today}\ }\textbf {\bibinfo {volume}
  {148}},\ \bibinfo {pages} {191} (\bibinfo {year} {2009})},\ \bibinfo {note}
  {special Issue of the 10th International Conference on CO$_2$ Utilization,
  Tianijn, China, May 17-21, 2009}\BibitemShut {NoStop}%
\bibitem [{\citenamefont {Compton}, \citenamefont {Reinhardt},\ and\
  \citenamefont {Cooper}(1975)}]{compton_jchemphys_1975}%
  \BibitemOpen
  \bibfield  {author} {\bibinfo {author} {\bibfnamefont {R.~N.}\ \bibnamefont
  {Compton}}, \bibinfo {author} {\bibfnamefont {P.~W.}\ \bibnamefont
  {Reinhardt}}, \ and\ \bibinfo {author} {\bibfnamefont {C.~D.}\ \bibnamefont
  {Cooper}},\ }\href {\doibase 10.1063/1.431875} {\bibfield  {journal}
  {\bibinfo  {journal} {The Journal of Chemical Physics}\ }\textbf {\bibinfo
  {volume} {63}},\ \bibinfo {pages} {3821} (\bibinfo {year}
  {1975})}\BibitemShut {NoStop}%
\bibitem [{\citenamefont {Benson}\ \emph {et~al.}(2009)\citenamefont {Benson},
  \citenamefont {Kubiak}, \citenamefont {Sathrum},\ and\ \citenamefont
  {Smieja}}]{benson_chemsocrev_2009}%
  \BibitemOpen
  \bibfield  {author} {\bibinfo {author} {\bibfnamefont {E.~E.}\ \bibnamefont
  {Benson}}, \bibinfo {author} {\bibfnamefont {C.~P.}\ \bibnamefont {Kubiak}},
  \bibinfo {author} {\bibfnamefont {A.~J.}\ \bibnamefont {Sathrum}}, \ and\
  \bibinfo {author} {\bibfnamefont {J.~M.}\ \bibnamefont {Smieja}},\ }\href
  {\doibase 10.1039/B804323J} {\bibfield  {journal} {\bibinfo  {journal}
  {Chemical Society Reviews}\ }\textbf {\bibinfo {volume} {38}},\ \bibinfo
  {pages} {89} (\bibinfo {year} {2009})}\BibitemShut {NoStop}%
\bibitem [{\citenamefont {Kohno}\ \emph {et~al.}(2000)\citenamefont {Kohno},
  \citenamefont {Tanaka}, \citenamefont {Funabiki},\ and\ \citenamefont
  {Yoshida}}]{kohno_physchemchemphys_2000}%
  \BibitemOpen
  \bibfield  {author} {\bibinfo {author} {\bibfnamefont {Y.}~\bibnamefont
  {Kohno}}, \bibinfo {author} {\bibfnamefont {T.}~\bibnamefont {Tanaka}},
  \bibinfo {author} {\bibfnamefont {T.}~\bibnamefont {Funabiki}}, \ and\
  \bibinfo {author} {\bibfnamefont {S.}~\bibnamefont {Yoshida}},\ }\href
  {\doibase 10.1039/B001642J} {\bibfield  {journal} {\bibinfo  {journal}
  {Physical Chemistry Chemical Physics}\ }\textbf {\bibinfo {volume} {2}},\
  \bibinfo {pages} {2635} (\bibinfo {year} {2000})}\BibitemShut {NoStop}%
\bibitem [{\citenamefont {Markovits}, \citenamefont {Fahmi},\ and\
  \citenamefont {Minot}(1996)}]{markovits_jmolstructheochem_1996}%
  \BibitemOpen
  \bibfield  {author} {\bibinfo {author} {\bibfnamefont {A.}~\bibnamefont
  {Markovits}}, \bibinfo {author} {\bibfnamefont {A.}~\bibnamefont {Fahmi}}, \
  and\ \bibinfo {author} {\bibfnamefont {C.}~\bibnamefont {Minot}},\ }\href
  {\doibase 10.1016/S0166-1280(96)04696-9} {\bibfield  {journal} {\bibinfo
  {journal} {Journal of Molecular Structure: THEOCHEM}\ }\textbf {\bibinfo
  {volume} {371}},\ \bibinfo {pages} {219 } (\bibinfo {year}
  {1996})}\BibitemShut {NoStop}%
\bibitem [{\citenamefont {He}, \citenamefont {Zapol},\ and\ \citenamefont
  {Curtiss}(2010)}]{he_jphyschemc_2010}%
  \BibitemOpen
  \bibfield  {author} {\bibinfo {author} {\bibfnamefont {H.}~\bibnamefont
  {He}}, \bibinfo {author} {\bibfnamefont {P.}~\bibnamefont {Zapol}}, \ and\
  \bibinfo {author} {\bibfnamefont {L.~A.}\ \bibnamefont {Curtiss}},\ }\href
  {\doibase 10.1021/jp106579b} {\bibfield  {journal} {\bibinfo  {journal} {The
  Journal of Physical Chemistry C}\ }\textbf {\bibinfo {volume} {114}},\
  \bibinfo {pages} {21474} (\bibinfo {year} {2010})},\ \Eprint
  {http://arxiv.org/abs/http://pubs.acs.org/doi/pdf/10.1021/jp106579b}
  {http://pubs.acs.org/doi/pdf/10.1021/jp106579b} \BibitemShut {NoStop}%
\bibitem [{\citenamefont {Sorescu}\ \emph {et~al.}(2011)\citenamefont
  {Sorescu}, \citenamefont {Lee}, \citenamefont {Al-Saidi},\ and\ \citenamefont
  {Jordan}}]{sorescu_jchemphys_2011a}%
  \BibitemOpen
  \bibfield  {author} {\bibinfo {author} {\bibfnamefont {D.~C.}\ \bibnamefont
  {Sorescu}}, \bibinfo {author} {\bibfnamefont {J.}~\bibnamefont {Lee}},
  \bibinfo {author} {\bibfnamefont {W.~A.}\ \bibnamefont {Al-Saidi}}, \ and\
  \bibinfo {author} {\bibfnamefont {K.~D.}\ \bibnamefont {Jordan}},\ }\href
  {\doibase 10.1063/1.3561300} {\bibfield  {journal} {\bibinfo  {journal} {The
  Journal of Chemical Physics}\ }\textbf {\bibinfo {volume} {134}},\ \bibinfo
  {eid} {104707} (\bibinfo {year} {2011})}\BibitemShut {NoStop}%
\bibitem [{\citenamefont {Sorescu}, \citenamefont {Al-Saidi},\ and\
  \citenamefont {Jordan}(2011)}]{sorescu_jchemphys_2011b}%
  \BibitemOpen
  \bibfield  {author} {\bibinfo {author} {\bibfnamefont {D.~C.}\ \bibnamefont
  {Sorescu}}, \bibinfo {author} {\bibfnamefont {W.~A.}\ \bibnamefont
  {Al-Saidi}}, \ and\ \bibinfo {author} {\bibfnamefont {K.~D.}\ \bibnamefont
  {Jordan}},\ }\href {\doibase 10.1063/1.3638181} {\bibfield  {journal}
  {\bibinfo  {journal} {The Journal of Chemical Physics}\ }\textbf {\bibinfo
  {volume} {135}},\ \bibinfo {eid} {124701} (\bibinfo {year}
  {2011})}\BibitemShut {NoStop}%
\bibitem [{\citenamefont {Teramura}\ \emph {et~al.}(2004)\citenamefont
  {Teramura}, \citenamefont {Tanaka}, \citenamefont {Ishikawa}, \citenamefont
  {Kohno},\ and\ \citenamefont {Funabiki}}]{teramura_jphyschemb_2004}%
  \BibitemOpen
  \bibfield  {author} {\bibinfo {author} {\bibfnamefont {K.}~\bibnamefont
  {Teramura}}, \bibinfo {author} {\bibfnamefont {T.}~\bibnamefont {Tanaka}},
  \bibinfo {author} {\bibfnamefont {H.}~\bibnamefont {Ishikawa}}, \bibinfo
  {author} {\bibfnamefont {Y.}~\bibnamefont {Kohno}}, \ and\ \bibinfo {author}
  {\bibfnamefont {T.}~\bibnamefont {Funabiki}},\ }\href {\doibase
  10.1021/jp0362943} {\bibfield  {journal} {\bibinfo  {journal} {The Journal of
  Physical Chemistry B}\ }\textbf {\bibinfo {volume} {108}},\ \bibinfo {pages}
  {346} (\bibinfo {year} {2004})},\ \Eprint
  {http://arxiv.org/abs/http://pubs.acs.org/doi/pdf/10.1021/jp0362943}
  {http://pubs.acs.org/doi/pdf/10.1021/jp0362943} \BibitemShut {NoStop}%
\bibitem [{\citenamefont {Besson}, \citenamefont {Vargas},\ and\ \citenamefont
  {Favergeon}(2012)}]{besson_surfsci_2012}%
  \BibitemOpen
  \bibfield  {author} {\bibinfo {author} {\bibfnamefont {R.}~\bibnamefont
  {Besson}}, \bibinfo {author} {\bibfnamefont {M.~R.}\ \bibnamefont {Vargas}},
  \ and\ \bibinfo {author} {\bibfnamefont {L.}~\bibnamefont {Favergeon}},\
  }\href {\doibase 10.1016/j.susc.2011.11.016} {\bibfield  {journal} {\bibinfo
  {journal} {Surface Science}\ }\textbf {\bibinfo {volume} {606}},\ \bibinfo
  {pages} {490} (\bibinfo {year} {2012})}\BibitemShut {NoStop}%
\bibitem [{\citenamefont {Trovarelli}(2002)}]{trovarelli_book_2002}%
  \BibitemOpen
  \bibfield  {author} {\bibinfo {author} {\bibfnamefont {A.}~\bibnamefont
  {Trovarelli}},\ }\href@noop {} {\emph {\bibinfo {title} {Catalysis by ceria
  and related materials}}},\ Vol.~\bibinfo {volume} {2}\ (\bibinfo  {publisher}
  {Imperial College Press},\ \bibinfo {address} {London},\ \bibinfo {year}
  {2002})\BibitemShut {NoStop}%
\bibitem [{\citenamefont {Kim}(1982)}]{kim_indengchemprodrd_1982}%
  \BibitemOpen
  \bibfield  {author} {\bibinfo {author} {\bibfnamefont {G.}~\bibnamefont
  {Kim}},\ }\href {\doibase 10.1021/i300006a014} {\bibfield  {journal}
  {\bibinfo  {journal} {Industrial \& Engineering Chemistry Product Research
  and Development}\ }\textbf {\bibinfo {volume} {21}},\ \bibinfo {pages} {267}
  (\bibinfo {year} {1982})},\ \Eprint
  {http://arxiv.org/abs/http://pubs.acs.org/doi/pdf/10.1021/i300006a014}
  {http://pubs.acs.org/doi/pdf/10.1021/i300006a014} \BibitemShut {NoStop}%
\bibitem [{\citenamefont {Bunluesin}, \citenamefont {Gorte},\ and\
  \citenamefont {Graham}(1998)}]{bunluesin_applcatalb_1998}%
  \BibitemOpen
  \bibfield  {author} {\bibinfo {author} {\bibfnamefont {T.}~\bibnamefont
  {Bunluesin}}, \bibinfo {author} {\bibfnamefont {R.~J.}\ \bibnamefont
  {Gorte}}, \ and\ \bibinfo {author} {\bibfnamefont {G.~W.}\ \bibnamefont
  {Graham}},\ }\href
  {http://www.sciencedirect.com/science/article/pii/S0926337397000404}
  {\bibfield  {journal} {\bibinfo  {journal} {Applied Catalysis B:
  Environmental}\ }\textbf {\bibinfo {volume} {15}},\ \bibinfo {pages} {107}
  (\bibinfo {year} {1998})}\BibitemShut {NoStop}%
\bibitem [{\citenamefont {Hilaire}\ \emph {et~al.}(2004)\citenamefont
  {Hilaire}, \citenamefont {Wang}, \citenamefont {Luo}, \citenamefont {Gorte},\
  and\ \citenamefont {Wagner}}]{hilaire_applcatala_2004}%
  \BibitemOpen
  \bibfield  {author} {\bibinfo {author} {\bibfnamefont {S.}~\bibnamefont
  {Hilaire}}, \bibinfo {author} {\bibfnamefont {X.}~\bibnamefont {Wang}},
  \bibinfo {author} {\bibfnamefont {T.}~\bibnamefont {Luo}}, \bibinfo {author}
  {\bibfnamefont {R.~J.}\ \bibnamefont {Gorte}}, \ and\ \bibinfo {author}
  {\bibfnamefont {J.}~\bibnamefont {Wagner}},\ }\href {\doibase DOI:
  10.1016/j.apcata.2003.09.026} {\bibfield  {journal} {\bibinfo  {journal}
  {Applied Catalysis A: General}\ }\textbf {\bibinfo {volume} {258}},\ \bibinfo
  {pages} {271} (\bibinfo {year} {2004})}\BibitemShut {NoStop}%
\bibitem [{\citenamefont {Pino}\ \emph {et~al.}(2002)\citenamefont {Pino},
  \citenamefont {Recupero}, \citenamefont {Beninati}, \citenamefont {Shukla},
  \citenamefont {Hegde},\ and\ \citenamefont {Bera}}]{pino_applcatala_2002}%
  \BibitemOpen
  \bibfield  {author} {\bibinfo {author} {\bibfnamefont {L.}~\bibnamefont
  {Pino}}, \bibinfo {author} {\bibfnamefont {V.}~\bibnamefont {Recupero}},
  \bibinfo {author} {\bibfnamefont {S.}~\bibnamefont {Beninati}}, \bibinfo
  {author} {\bibfnamefont {A.~K.}\ \bibnamefont {Shukla}}, \bibinfo {author}
  {\bibfnamefont {M.~S.}\ \bibnamefont {Hegde}}, \ and\ \bibinfo {author}
  {\bibfnamefont {P.}~\bibnamefont {Bera}},\ }\href
  {http://www.sciencedirect.com/science/article/pii/S0926860X01007347}
  {\bibfield  {journal} {\bibinfo  {journal} {Applied Catalysis A: General}\
  }\textbf {\bibinfo {volume} {225}},\ \bibinfo {pages} {63} (\bibinfo {year}
  {2002})}\BibitemShut {NoStop}%
\bibitem [{\citenamefont {Pino}\ \emph {et~al.}(2003)\citenamefont {Pino},
  \citenamefont {Vita}, \citenamefont {Cordaro}, \citenamefont {Recupero},\
  and\ \citenamefont {Hegde}}]{pino_applcatala_2003}%
  \BibitemOpen
  \bibfield  {author} {\bibinfo {author} {\bibfnamefont {L.}~\bibnamefont
  {Pino}}, \bibinfo {author} {\bibfnamefont {A.}~\bibnamefont {Vita}}, \bibinfo
  {author} {\bibfnamefont {M.}~\bibnamefont {Cordaro}}, \bibinfo {author}
  {\bibfnamefont {V.}~\bibnamefont {Recupero}}, \ and\ \bibinfo {author}
  {\bibfnamefont {M.}~\bibnamefont {Hegde}},\ }\href {\doibase
  doi:10.1016/S0926-860X(02)00542-2} {\bibfield  {journal} {\bibinfo  {journal}
  {Applied Catalysis A: General}\ }\textbf {\bibinfo {volume} {243}},\ \bibinfo
  {pages} {135} (\bibinfo {year} {2003})}\BibitemShut {NoStop}%
\bibitem [{\citenamefont {Herschend}, \citenamefont {Baudin},\ and\
  \citenamefont {Hermansson}(2006)}]{herschend_chemphys_2006}%
  \BibitemOpen
  \bibfield  {author} {\bibinfo {author} {\bibfnamefont {B.}~\bibnamefont
  {Herschend}}, \bibinfo {author} {\bibfnamefont {M.}~\bibnamefont {Baudin}}, \
  and\ \bibinfo {author} {\bibfnamefont {K.}~\bibnamefont {Hermansson}},\
  }\href {\doibase 10.1016/j.chemphys.2006.07.022} {\bibfield  {journal}
  {\bibinfo  {journal} {Chemical Physics}\ }\textbf {\bibinfo {volume} {328}},\
  \bibinfo {pages} {345} (\bibinfo {year} {2006})}\BibitemShut {NoStop}%
\bibitem [{\citenamefont {Yang}, \citenamefont {Woo},\ and\ \citenamefont
  {Hermansson}(2004)}]{yang_chemphyslett_2004}%
  \BibitemOpen
  \bibfield  {author} {\bibinfo {author} {\bibfnamefont {Z.}~\bibnamefont
  {Yang}}, \bibinfo {author} {\bibfnamefont {T.~K.}\ \bibnamefont {Woo}}, \
  and\ \bibinfo {author} {\bibfnamefont {K.}~\bibnamefont {Hermansson}},\
  }\href {\doibase 10.1016/j.cplett.2004.08.078} {\bibfield  {journal}
  {\bibinfo  {journal} {Chemical Physics Letters}\ }\textbf {\bibinfo {volume}
  {396}},\ \bibinfo {pages} {384} (\bibinfo {year} {2004})}\BibitemShut
  {NoStop}%
\bibitem [{\citenamefont {Bernal}\ \emph {et~al.}(2001)\citenamefont {Bernal},
  \citenamefont {Blanco}, \citenamefont {Gatica}, \citenamefont {Larese},\ and\
  \citenamefont {Vidal}}]{bernal_jcatal_2001}%
  \BibitemOpen
  \bibfield  {author} {\bibinfo {author} {\bibfnamefont {S.}~\bibnamefont
  {Bernal}}, \bibinfo {author} {\bibfnamefont {G.}~\bibnamefont {Blanco}},
  \bibinfo {author} {\bibfnamefont {J.}~\bibnamefont {Gatica}}, \bibinfo
  {author} {\bibfnamefont {C.}~\bibnamefont {Larese}}, \ and\ \bibinfo {author}
  {\bibfnamefont {H.}~\bibnamefont {Vidal}},\ }\href {\doibase
  10.1006/jcat.2001.3210} {\bibfield  {journal} {\bibinfo  {journal} {Journal
  of Catalysis}\ }\textbf {\bibinfo {volume} {200}},\ \bibinfo {pages} {411 }
  (\bibinfo {year} {2001})}\BibitemShut {NoStop}%
\bibitem [{\citenamefont {Jin}\ \emph {et~al.}(1987)\citenamefont {Jin},
  \citenamefont {Zhou}, \citenamefont {Mains},\ and\ \citenamefont
  {White}}]{jin_jphyschem_1987}%
  \BibitemOpen
  \bibfield  {author} {\bibinfo {author} {\bibfnamefont {T.}~\bibnamefont
  {Jin}}, \bibinfo {author} {\bibfnamefont {Y.}~\bibnamefont {Zhou}}, \bibinfo
  {author} {\bibfnamefont {G.~J.}\ \bibnamefont {Mains}}, \ and\ \bibinfo
  {author} {\bibfnamefont {J.~M.}\ \bibnamefont {White}},\ }\href {\doibase
  10.1021/j100307a023} {\bibfield  {journal} {\bibinfo  {journal} {The Journal
  of Physical Chemistry}\ }\textbf {\bibinfo {volume} {91}},\ \bibinfo {pages}
  {5931} (\bibinfo {year} {1987})},\ \Eprint
  {http://arxiv.org/abs/http://pubs.acs.org/doi/pdf/10.1021/j100307a023}
  {http://pubs.acs.org/doi/pdf/10.1021/j100307a023} \BibitemShut {NoStop}%
\bibitem [{\citenamefont {Otsuka}\ \emph {et~al.}(1998)\citenamefont {Otsuka},
  \citenamefont {Wang}, \citenamefont {Sunada},\ and\ \citenamefont
  {Yamanaka}}]{otsuka_jcatal_1998}%
  \BibitemOpen
  \bibfield  {author} {\bibinfo {author} {\bibfnamefont {K.}~\bibnamefont
  {Otsuka}}, \bibinfo {author} {\bibfnamefont {Y.}~\bibnamefont {Wang}},
  \bibinfo {author} {\bibfnamefont {E.}~\bibnamefont {Sunada}}, \ and\ \bibinfo
  {author} {\bibfnamefont {I.}~\bibnamefont {Yamanaka}},\ }\href {\doibase
  10.1006/jcat.1998.1985} {\bibfield  {journal} {\bibinfo  {journal} {Journal
  of Catalysis}\ }\textbf {\bibinfo {volume} {175}},\ \bibinfo {pages} {152 }
  (\bibinfo {year} {1998})}\BibitemShut {NoStop}%
\bibitem [{\citenamefont {Bueno-L{\'o}pez}, \citenamefont {Krishna},\ and\
  \citenamefont {Makkee}(2008)}]{buenolopez_applcatalagen_2008}%
  \BibitemOpen
  \bibfield  {author} {\bibinfo {author} {\bibfnamefont {A.}~\bibnamefont
  {Bueno-L{\'o}pez}}, \bibinfo {author} {\bibfnamefont {K.}~\bibnamefont
  {Krishna}}, \ and\ \bibinfo {author} {\bibfnamefont {M.}~\bibnamefont
  {Makkee}},\ }\href
  {http://www.sciencedirect.com/science/article/pii/S0926860X08001725}
  {\bibfield  {journal} {\bibinfo  {journal} {Applied Catalysis A: General}\
  }\textbf {\bibinfo {volume} {342}},\ \bibinfo {pages} {144} (\bibinfo {year}
  {2008})}\BibitemShut {NoStop}%
\bibitem [{\citenamefont {Staudt}\ \emph {et~al.}(2010)\citenamefont {Staudt},
  \citenamefont {Lykhach}, \citenamefont {Tsud}, \citenamefont {Sk{\'a}la},
  \citenamefont {Prince}, \citenamefont {Matol{\'\i}n},\ and\ \citenamefont
  {Libuda}}]{staudt_jcatal_2010}%
  \BibitemOpen
  \bibfield  {author} {\bibinfo {author} {\bibfnamefont {T.}~\bibnamefont
  {Staudt}}, \bibinfo {author} {\bibfnamefont {Y.}~\bibnamefont {Lykhach}},
  \bibinfo {author} {\bibfnamefont {N.}~\bibnamefont {Tsud}}, \bibinfo {author}
  {\bibfnamefont {T.}~\bibnamefont {Sk{\'a}la}}, \bibinfo {author}
  {\bibfnamefont {K.}~\bibnamefont {Prince}}, \bibinfo {author} {\bibfnamefont
  {V.}~\bibnamefont {Matol{\'\i}n}}, \ and\ \bibinfo {author} {\bibfnamefont
  {J.}~\bibnamefont {Libuda}},\ }\href {\doibase 10.1016/j.jcat.2010.07.032}
  {\bibfield  {journal} {\bibinfo  {journal} {Journal of Catalysis}\ }\textbf
  {\bibinfo {volume} {275}},\ \bibinfo {pages} {181} (\bibinfo {year}
  {2010})}\BibitemShut {NoStop}%
\bibitem [{\citenamefont {Yang}\ \emph {et~al.}(2004)\citenamefont {Yang},
  \citenamefont {Woo}, \citenamefont {Baudin},\ and\ \citenamefont
  {Hermansson}}]{yang_jchemphys_2004}%
  \BibitemOpen
  \bibfield  {author} {\bibinfo {author} {\bibfnamefont {Z.}~\bibnamefont
  {Yang}}, \bibinfo {author} {\bibfnamefont {T.~K.}\ \bibnamefont {Woo}},
  \bibinfo {author} {\bibfnamefont {M.}~\bibnamefont {Baudin}}, \ and\ \bibinfo
  {author} {\bibfnamefont {K.}~\bibnamefont {Hermansson}},\ }\href {\doibase
  10.1063/1.1688316} {\bibfield  {journal} {\bibinfo  {journal} {The Journal of
  Chemical Physics}\ }\textbf {\bibinfo {volume} {120}},\ \bibinfo {pages}
  {7741} (\bibinfo {year} {2004})}\BibitemShut {NoStop}%
\bibitem [{\citenamefont {Fabris}\ \emph
  {et~al.}(2005{\natexlab{a}})\citenamefont {Fabris}, \citenamefont
  {de~Gironcoli}, \citenamefont {Baroni}, \citenamefont {Vicario},\ and\
  \citenamefont {Balducci}}]{fabris_physrevb_2005a}%
  \BibitemOpen
  \bibfield  {author} {\bibinfo {author} {\bibfnamefont {S.}~\bibnamefont
  {Fabris}}, \bibinfo {author} {\bibfnamefont {S.}~\bibnamefont
  {de~Gironcoli}}, \bibinfo {author} {\bibfnamefont {S.}~\bibnamefont
  {Baroni}}, \bibinfo {author} {\bibfnamefont {G.}~\bibnamefont {Vicario}}, \
  and\ \bibinfo {author} {\bibfnamefont {G.}~\bibnamefont {Balducci}},\ }\href
  {\doibase 10.1103/PhysRevB.71.041102} {\bibfield  {journal} {\bibinfo
  {journal} {Physical Review B}\ }\textbf {\bibinfo {volume} {71}},\ \bibinfo
  {pages} {041102} (\bibinfo {year} {2005}{\natexlab{a}})}\BibitemShut
  {NoStop}%
\bibitem [{\citenamefont {Burbano}\ \emph {et~al.}(2011)\citenamefont
  {Burbano}, \citenamefont {Marrocchelli}, \citenamefont {Yildiz},
  \citenamefont {Tuller}, \citenamefont {Norberg}, \citenamefont {Hull},
  \citenamefont {Madden},\ and\ \citenamefont
  {Watson}}]{burbano_jphyscondensmat_2011}%
  \BibitemOpen
  \bibfield  {author} {\bibinfo {author} {\bibfnamefont {M.}~\bibnamefont
  {Burbano}}, \bibinfo {author} {\bibfnamefont {D.}~\bibnamefont
  {Marrocchelli}}, \bibinfo {author} {\bibfnamefont {B.}~\bibnamefont
  {Yildiz}}, \bibinfo {author} {\bibfnamefont {H.~L.}\ \bibnamefont {Tuller}},
  \bibinfo {author} {\bibfnamefont {S.~T.}\ \bibnamefont {Norberg}}, \bibinfo
  {author} {\bibfnamefont {S.}~\bibnamefont {Hull}}, \bibinfo {author}
  {\bibfnamefont {P.~A.}\ \bibnamefont {Madden}}, \ and\ \bibinfo {author}
  {\bibfnamefont {G.~W.}\ \bibnamefont {Watson}},\ }\href
  {http://stacks.iop.org/0953-8984/23/i=25/a=255402} {\bibfield  {journal}
  {\bibinfo  {journal} {Journal of Physics: Condensed Matter}\ }\textbf
  {\bibinfo {volume} {23}},\ \bibinfo {pages} {255402} (\bibinfo {year}
  {2011})}\BibitemShut {NoStop}%
\bibitem [{\citenamefont {Skorodumova}, \citenamefont {Baudin},\ and\
  \citenamefont {Hermansson}(2004)}]{skorodumova_physrevb_2004}%
  \BibitemOpen
  \bibfield  {author} {\bibinfo {author} {\bibfnamefont {N.~V.}\ \bibnamefont
  {Skorodumova}}, \bibinfo {author} {\bibfnamefont {M.}~\bibnamefont {Baudin}},
  \ and\ \bibinfo {author} {\bibfnamefont {K.}~\bibnamefont {Hermansson}},\
  }\href {\doibase 10.1103/PhysRevB.69.075401} {\bibfield  {journal} {\bibinfo
  {journal} {Physical Review B}\ }\textbf {\bibinfo {volume} {69}},\ \bibinfo
  {pages} {075401} (\bibinfo {year} {2004})}\BibitemShut {NoStop}%
\bibitem [{\citenamefont {Herschend}, \citenamefont {Baudin},\ and\
  \citenamefont {Hermansson}(2005)}]{herschend_surfsci_2005}%
  \BibitemOpen
  \bibfield  {author} {\bibinfo {author} {\bibfnamefont {B.}~\bibnamefont
  {Herschend}}, \bibinfo {author} {\bibfnamefont {M.}~\bibnamefont {Baudin}}, \
  and\ \bibinfo {author} {\bibfnamefont {K.}~\bibnamefont {Hermansson}},\
  }\href {\doibase 10.1016/j.susc.2005.09.045} {\bibfield  {journal} {\bibinfo
  {journal} {Surface Science}\ }\textbf {\bibinfo {volume} {599}},\ \bibinfo
  {pages} {173} (\bibinfo {year} {2005})}\BibitemShut {NoStop}%
\bibitem [{\citenamefont {Ganduglia-Pirovano}, \citenamefont {Da~Silva},\ and\
  \citenamefont {Sauer}(2009)}]{gandugliapirovano_physrevlett_2009}%
  \BibitemOpen
  \bibfield  {author} {\bibinfo {author} {\bibfnamefont {M.~V.}\ \bibnamefont
  {Ganduglia-Pirovano}}, \bibinfo {author} {\bibfnamefont {J.~L.~F.}\
  \bibnamefont {Da~Silva}}, \ and\ \bibinfo {author} {\bibfnamefont
  {J.}~\bibnamefont {Sauer}},\ }\href {\doibase 10.1103/PhysRevLett.102.026101}
  {\bibfield  {journal} {\bibinfo  {journal} {Physical Review Letters}\
  }\textbf {\bibinfo {volume} {102}},\ \bibinfo {pages} {026101} (\bibinfo
  {year} {2009})}\BibitemShut {NoStop}%
\bibitem [{\citenamefont {Yang}\ \emph {et~al.}(2009)\citenamefont {Yang},
  \citenamefont {Yu}, \citenamefont {Lu}, \citenamefont {Li},\ and\
  \citenamefont {Hermansson}}]{yang_physletta_2009}%
  \BibitemOpen
  \bibfield  {author} {\bibinfo {author} {\bibfnamefont {Z.}~\bibnamefont
  {Yang}}, \bibinfo {author} {\bibfnamefont {X.}~\bibnamefont {Yu}}, \bibinfo
  {author} {\bibfnamefont {Z.}~\bibnamefont {Lu}}, \bibinfo {author}
  {\bibfnamefont {S.}~\bibnamefont {Li}}, \ and\ \bibinfo {author}
  {\bibfnamefont {K.}~\bibnamefont {Hermansson}},\ }\href {\doibase
  10.1016/j.physleta.2009.05.055} {\bibfield  {journal} {\bibinfo  {journal}
  {Physics Letters A}\ }\textbf {\bibinfo {volume} {373}},\ \bibinfo {pages}
  {2786 } (\bibinfo {year} {2009})}\BibitemShut {NoStop}%
\bibitem [{\citenamefont {Nolan}\ \emph {et~al.}(2005)\citenamefont {Nolan},
  \citenamefont {Grigoleit}, \citenamefont {Sayle}, \citenamefont {Parker},\
  and\ \citenamefont {Watson}}]{nolan_surfsci_2005a}%
  \BibitemOpen
  \bibfield  {author} {\bibinfo {author} {\bibfnamefont {M.}~\bibnamefont
  {Nolan}}, \bibinfo {author} {\bibfnamefont {S.}~\bibnamefont {Grigoleit}},
  \bibinfo {author} {\bibfnamefont {D.~C.}\ \bibnamefont {Sayle}}, \bibinfo
  {author} {\bibfnamefont {S.~C.}\ \bibnamefont {Parker}}, \ and\ \bibinfo
  {author} {\bibfnamefont {G.~W.}\ \bibnamefont {Watson}},\ }\href {\doibase
  10.1016/j.susc.2004.12.016} {\bibfield  {journal} {\bibinfo  {journal}
  {Surface Science}\ }\textbf {\bibinfo {volume} {576}},\ \bibinfo {pages}
  {217} (\bibinfo {year} {2005})}\BibitemShut {NoStop}%
\bibitem [{\citenamefont {Nolan}, \citenamefont {Parker},\ and\ \citenamefont
  {Watson}(2005)}]{nolan_surfsci_2005b}%
  \BibitemOpen
  \bibfield  {author} {\bibinfo {author} {\bibfnamefont {M.}~\bibnamefont
  {Nolan}}, \bibinfo {author} {\bibfnamefont {S.~C.}\ \bibnamefont {Parker}}, \
  and\ \bibinfo {author} {\bibfnamefont {G.~W.}\ \bibnamefont {Watson}},\
  }\href@noop {} {\bibfield  {journal} {\bibinfo  {journal} {Surface Science}\
  }\textbf {\bibinfo {volume} {595}},\ \bibinfo {pages} {223} (\bibinfo {year}
  {2005})}\BibitemShut {NoStop}%
\bibitem [{\citenamefont {Hohenberg}\ and\ \citenamefont
  {Kohn}(1964)}]{hohenberg_physrev_1964}%
  \BibitemOpen
  \bibfield  {author} {\bibinfo {author} {\bibfnamefont {P.}~\bibnamefont
  {Hohenberg}}\ and\ \bibinfo {author} {\bibfnamefont {W.}~\bibnamefont
  {Kohn}},\ }\href {\doibase 10.1103/PhysRev.136.B864} {\bibfield  {journal}
  {\bibinfo  {journal} {Physical Review}\ }\textbf {\bibinfo {volume} {136}},\
  \bibinfo {pages} {B864} (\bibinfo {year} {1964})}\BibitemShut {NoStop}%
\bibitem [{\citenamefont {Kohn}\ and\ \citenamefont
  {Sham}(1965)}]{kohn_physrev_1965}%
  \BibitemOpen
  \bibfield  {author} {\bibinfo {author} {\bibfnamefont {W.}~\bibnamefont
  {Kohn}}\ and\ \bibinfo {author} {\bibfnamefont {L.~J.}\ \bibnamefont
  {Sham}},\ }\href {\doibase 10.1103/PhysRev.140.A1133} {\bibfield  {journal}
  {\bibinfo  {journal} {Physical Review}\ }\textbf {\bibinfo {volume} {140}},\
  \bibinfo {pages} {A1133} (\bibinfo {year} {1965})}\BibitemShut {NoStop}%
\bibitem [{\citenamefont {Kresse}\ and\ \citenamefont
  {Hafner}(1993)}]{kresse_physrevb_1993}%
  \BibitemOpen
  \bibfield  {author} {\bibinfo {author} {\bibfnamefont {G.}~\bibnamefont
  {Kresse}}\ and\ \bibinfo {author} {\bibfnamefont {J.}~\bibnamefont
  {Hafner}},\ }\href {\doibase 10.1103/PhysRevB.47.558} {\bibfield  {journal}
  {\bibinfo  {journal} {Physical Review B}\ }\textbf {\bibinfo {volume} {47}},\
  \bibinfo {pages} {558} (\bibinfo {year} {1993})}\BibitemShut {NoStop}%
\bibitem [{\citenamefont {Kresse}\ and\ \citenamefont
  {Furthm{\"u}ller}(1996{\natexlab{a}})}]{kresse_compmatersci_1996}%
  \BibitemOpen
  \bibfield  {author} {\bibinfo {author} {\bibfnamefont {G.}~\bibnamefont
  {Kresse}}\ and\ \bibinfo {author} {\bibfnamefont {J.}~\bibnamefont
  {Furthm{\"u}ller}},\ }\href {\doibase 10.1016/0927-0256(96)00008-0}
  {\bibfield  {journal} {\bibinfo  {journal} {Computational Materials Science}\
  }\textbf {\bibinfo {volume} {6}},\ \bibinfo {pages} {15} (\bibinfo {year}
  {1996}{\natexlab{a}})}\BibitemShut {NoStop}%
\bibitem [{\citenamefont {Kresse}\ and\ \citenamefont
  {Furthm{\"u}ller}(1996{\natexlab{b}})}]{kresse_physrevb_1996}%
  \BibitemOpen
  \bibfield  {author} {\bibinfo {author} {\bibfnamefont {G.}~\bibnamefont
  {Kresse}}\ and\ \bibinfo {author} {\bibfnamefont {J.}~\bibnamefont
  {Furthm{\"u}ller}},\ }\href {\doibase 10.1103/PhysRevB.54.11169} {\bibfield
  {journal} {\bibinfo  {journal} {Physical Review B}\ }\textbf {\bibinfo
  {volume} {54}},\ \bibinfo {pages} {11169} (\bibinfo {year}
  {1996}{\natexlab{b}})}\BibitemShut {NoStop}%
\bibitem [{\citenamefont {Perdew}, \citenamefont {Burke},\ and\ \citenamefont
  {Ernzerhof}(1996)}]{perdew_physrevlett_1996}%
  \BibitemOpen
  \bibfield  {author} {\bibinfo {author} {\bibfnamefont {J.~P.}\ \bibnamefont
  {Perdew}}, \bibinfo {author} {\bibfnamefont {K.}~\bibnamefont {Burke}}, \
  and\ \bibinfo {author} {\bibfnamefont {M.}~\bibnamefont {Ernzerhof}},\ }\href
  {\doibase 10.1103/PhysRevLett.77.3865} {\bibfield  {journal} {\bibinfo
  {journal} {Physical Review Letters}\ }\textbf {\bibinfo {volume} {77}},\
  \bibinfo {pages} {3865} (\bibinfo {year} {1996})}\BibitemShut {NoStop}%
\bibitem [{\citenamefont {Bl{\"o}chl}(1994)}]{blochl_physrevb_1994b}%
  \BibitemOpen
  \bibfield  {author} {\bibinfo {author} {\bibfnamefont {P.~E.}\ \bibnamefont
  {Bl{\"o}chl}},\ }\href {\doibase 10.1103/PhysRevB.50.17953} {\bibfield
  {journal} {\bibinfo  {journal} {Physical Review B}\ }\textbf {\bibinfo
  {volume} {50}},\ \bibinfo {pages} {17953} (\bibinfo {year}
  {1994})}\BibitemShut {NoStop}%
\bibitem [{\citenamefont {Kresse}\ and\ \citenamefont
  {Joubert}(1999)}]{kresse_physrevb_1999}%
  \BibitemOpen
  \bibfield  {author} {\bibinfo {author} {\bibfnamefont {G.}~\bibnamefont
  {Kresse}}\ and\ \bibinfo {author} {\bibfnamefont {D.}~\bibnamefont
  {Joubert}},\ }\href@noop {} {\bibfield  {journal} {\bibinfo  {journal}
  {Physical Review B}\ }\textbf {\bibinfo {volume} {59}},\ \bibinfo {pages}
  {1758} (\bibinfo {year} {1999})}\BibitemShut {NoStop}%
\bibitem [{\citenamefont {Bl{\"o}chl}, \citenamefont {Jepsen},\ and\
  \citenamefont {Andersen}(1994)}]{blochl_physrevb_1994a}%
  \BibitemOpen
  \bibfield  {author} {\bibinfo {author} {\bibfnamefont {P.~E.}\ \bibnamefont
  {Bl{\"o}chl}}, \bibinfo {author} {\bibfnamefont {O.}~\bibnamefont {Jepsen}},
  \ and\ \bibinfo {author} {\bibfnamefont {O.~K.}\ \bibnamefont {Andersen}},\
  }\href {\doibase 10.1103/PhysRevB.49.16223} {\bibfield  {journal} {\bibinfo
  {journal} {Physical Review B}\ }\textbf {\bibinfo {volume} {49}},\ \bibinfo
  {pages} {16223} (\bibinfo {year} {1994})}\BibitemShut {NoStop}%
\bibitem [{\citenamefont {K{\"u}mmerle}\ and\ \citenamefont
  {Heger}(1999)}]{kummerle_jsolidstatechem_1999}%
  \BibitemOpen
  \bibfield  {author} {\bibinfo {author} {\bibfnamefont {E.~A.}\ \bibnamefont
  {K{\"u}mmerle}}\ and\ \bibinfo {author} {\bibfnamefont {G.}~\bibnamefont
  {Heger}},\ }\href {\doibase 10.1006/jssc.1999.8403} {\bibfield  {journal}
  {\bibinfo  {journal} {Journal of Solid State Chemistry}\ }\textbf {\bibinfo
  {volume} {147}},\ \bibinfo {pages} {485} (\bibinfo {year}
  {1999})}\BibitemShut {NoStop}%
\bibitem [{\citenamefont {Skorodumova}\ \emph {et~al.}(2001)\citenamefont
  {Skorodumova}, \citenamefont {Ahuja}, \citenamefont {Simak}, \citenamefont
  {Abrikosov}, \citenamefont {Johansson},\ and\ \citenamefont
  {Lundqvist}}]{skorodumova_physrevb_2001}%
  \BibitemOpen
  \bibfield  {author} {\bibinfo {author} {\bibfnamefont {N.~V.}\ \bibnamefont
  {Skorodumova}}, \bibinfo {author} {\bibfnamefont {R.}~\bibnamefont {Ahuja}},
  \bibinfo {author} {\bibfnamefont {S.~I.}\ \bibnamefont {Simak}}, \bibinfo
  {author} {\bibfnamefont {I.~A.}\ \bibnamefont {Abrikosov}}, \bibinfo {author}
  {\bibfnamefont {B.}~\bibnamefont {Johansson}}, \ and\ \bibinfo {author}
  {\bibfnamefont {B.~I.}\ \bibnamefont {Lundqvist}},\ }\href {\doibase
  10.1103/PhysRevB.64.115108} {\bibfield  {journal} {\bibinfo  {journal}
  {Physical Review B}\ }\textbf {\bibinfo {volume} {64}},\ \bibinfo {pages}
  {115108} (\bibinfo {year} {2001})}\BibitemShut {NoStop}%
\bibitem [{\citenamefont {Koelling}, \citenamefont {Boring},\ and\
  \citenamefont {Wood}(1983)}]{koelling_solidstatecommun_1983}%
  \BibitemOpen
  \bibfield  {author} {\bibinfo {author} {\bibfnamefont {D.}~\bibnamefont
  {Koelling}}, \bibinfo {author} {\bibfnamefont {A.}~\bibnamefont {Boring}}, \
  and\ \bibinfo {author} {\bibfnamefont {J.}~\bibnamefont {Wood}},\ }\href
  {\doibase 10.1016/0038-1098(83)90550-1} {\bibfield  {journal} {\bibinfo
  {journal} {Solid State Communications}\ }\textbf {\bibinfo {volume} {47}},\
  \bibinfo {pages} {227} (\bibinfo {year} {1983})}\BibitemShut {NoStop}%
\bibitem [{\citenamefont {Hill}\ and\ \citenamefont
  {Catlow}(1993)}]{hill_jphyschemsolids_1993}%
  \BibitemOpen
  \bibfield  {author} {\bibinfo {author} {\bibfnamefont {S.}~\bibnamefont
  {Hill}}\ and\ \bibinfo {author} {\bibfnamefont {C.}~\bibnamefont {Catlow}},\
  }\href {\doibase 10.1016/0022-3697(93)90322-I} {\bibfield  {journal}
  {\bibinfo  {journal} {Journal of Physics and Chemistry of Solids}\ }\textbf
  {\bibinfo {volume} {54}},\ \bibinfo {pages} {411} (\bibinfo {year}
  {1993})}\BibitemShut {NoStop}%
\bibitem [{\citenamefont {Mullins}, \citenamefont {Overbury},\ and\
  \citenamefont {Huntley}(1998)}]{mullins_surfsci_1998}%
  \BibitemOpen
  \bibfield  {author} {\bibinfo {author} {\bibfnamefont {D.}~\bibnamefont
  {Mullins}}, \bibinfo {author} {\bibfnamefont {S.}~\bibnamefont {Overbury}}, \
  and\ \bibinfo {author} {\bibfnamefont {D.}~\bibnamefont {Huntley}},\ }\href
  {\doibase 10.1016/S0039-6028(98)00257-X} {\bibfield  {journal} {\bibinfo
  {journal} {Surface Science}\ }\textbf {\bibinfo {volume} {409}},\ \bibinfo
  {pages} {307} (\bibinfo {year} {1998})}\BibitemShut {NoStop}%
\bibitem [{\citenamefont {Kresse}\ \emph {et~al.}(2005)\citenamefont {Kresse},
  \citenamefont {Blaha}, \citenamefont {Da~Silva},\ and\ \citenamefont
  {Ganduglia-Pirovano}}]{kresse_physrevb_2005}%
  \BibitemOpen
  \bibfield  {author} {\bibinfo {author} {\bibfnamefont {G.}~\bibnamefont
  {Kresse}}, \bibinfo {author} {\bibfnamefont {P.}~\bibnamefont {Blaha}},
  \bibinfo {author} {\bibfnamefont {J.~L.~F.}\ \bibnamefont {Da~Silva}}, \ and\
  \bibinfo {author} {\bibfnamefont {M.~V.}\ \bibnamefont
  {Ganduglia-Pirovano}},\ }\href {\doibase 10.1103/PhysRevB.72.237101}
  {\bibfield  {journal} {\bibinfo  {journal} {Physical Review B}\ }\textbf
  {\bibinfo {volume} {72}},\ \bibinfo {pages} {237101} (\bibinfo {year}
  {2005})}\BibitemShut {NoStop}%
\bibitem [{\citenamefont {Fabris}\ \emph
  {et~al.}(2005{\natexlab{b}})\citenamefont {Fabris}, \citenamefont
  {de~Gironcoli}, \citenamefont {Baroni}, \citenamefont {Vicario},\ and\
  \citenamefont {Balducci}}]{fabris_physrevb_2005b}%
  \BibitemOpen
  \bibfield  {author} {\bibinfo {author} {\bibfnamefont {S.}~\bibnamefont
  {Fabris}}, \bibinfo {author} {\bibfnamefont {S.}~\bibnamefont
  {de~Gironcoli}}, \bibinfo {author} {\bibfnamefont {S.}~\bibnamefont
  {Baroni}}, \bibinfo {author} {\bibfnamefont {G.}~\bibnamefont {Vicario}}, \
  and\ \bibinfo {author} {\bibfnamefont {G.}~\bibnamefont {Balducci}},\ }\href
  {\doibase 10.1103/PhysRevB.72.237102} {\bibfield  {journal} {\bibinfo
  {journal} {Physical Review B}\ }\textbf {\bibinfo {volume} {72}},\ \bibinfo
  {pages} {237102} (\bibinfo {year} {2005}{\natexlab{b}})}\BibitemShut
  {NoStop}%
\bibitem [{\citenamefont {Herbst}, \citenamefont {Watson},\ and\ \citenamefont
  {Wilkins}(1978)}]{herbst_physrevb_1978}%
  \BibitemOpen
  \bibfield  {author} {\bibinfo {author} {\bibfnamefont {J.~F.}\ \bibnamefont
  {Herbst}}, \bibinfo {author} {\bibfnamefont {R.~E.}\ \bibnamefont {Watson}},
  \ and\ \bibinfo {author} {\bibfnamefont {J.~W.}\ \bibnamefont {Wilkins}},\
  }\href {\doibase 10.1103/PhysRevB.17.3089} {\bibfield  {journal} {\bibinfo
  {journal} {Physical Review B}\ }\textbf {\bibinfo {volume} {17}},\ \bibinfo
  {pages} {3089} (\bibinfo {year} {1978})}\BibitemShut {NoStop}%
\bibitem [{\citenamefont {Anisimov}\ and\ \citenamefont
  {Gunnarsson}(1991)}]{anisimov_physrevb_1991a}%
  \BibitemOpen
  \bibfield  {author} {\bibinfo {author} {\bibfnamefont {V.~I.}\ \bibnamefont
  {Anisimov}}\ and\ \bibinfo {author} {\bibfnamefont {O.}~\bibnamefont
  {Gunnarsson}},\ }\href {\doibase 10.1103/PhysRevB.43.7570} {\bibfield
  {journal} {\bibinfo  {journal} {Physical Review B}\ }\textbf {\bibinfo
  {volume} {43}},\ \bibinfo {pages} {7570} (\bibinfo {year}
  {1991})}\BibitemShut {NoStop}%
\bibitem [{\citenamefont {Loschen}\ \emph {et~al.}(2007)\citenamefont
  {Loschen}, \citenamefont {Carrasco}, \citenamefont {Neyman},\ and\
  \citenamefont {Illas}}]{loschen_physrevb_2007}%
  \BibitemOpen
  \bibfield  {author} {\bibinfo {author} {\bibfnamefont {C.}~\bibnamefont
  {Loschen}}, \bibinfo {author} {\bibfnamefont {J.}~\bibnamefont {Carrasco}},
  \bibinfo {author} {\bibfnamefont {K.~M.}\ \bibnamefont {Neyman}}, \ and\
  \bibinfo {author} {\bibfnamefont {F.}~\bibnamefont {Illas}},\ }\href
  {\doibase 10.1103/PhysRevB.75.035115} {\bibfield  {journal} {\bibinfo
  {journal} {Physical Review B}\ }\textbf {\bibinfo {volume} {75}},\ \bibinfo
  {pages} {035115} (\bibinfo {year} {2007})}\BibitemShut {NoStop}%
\bibitem [{\citenamefont {Sanville}\ \emph {et~al.}(2007)\citenamefont
  {Sanville}, \citenamefont {Kenny}, \citenamefont {Smith},\ and\ \citenamefont
  {Henkelman}}]{sanville_jcomputchem_2007}%
  \BibitemOpen
  \bibfield  {author} {\bibinfo {author} {\bibfnamefont {E.}~\bibnamefont
  {Sanville}}, \bibinfo {author} {\bibfnamefont {S.~D.}\ \bibnamefont {Kenny}},
  \bibinfo {author} {\bibfnamefont {R.}~\bibnamefont {Smith}}, \ and\ \bibinfo
  {author} {\bibfnamefont {G.}~\bibnamefont {Henkelman}},\ }\href@noop {}
  {\bibfield  {journal} {\bibinfo  {journal} {Journal of Computational
  Chemistry}\ }\textbf {\bibinfo {volume} {28}},\ \bibinfo {pages} {899}
  (\bibinfo {year} {2007})}\BibitemShut {NoStop}%
\bibitem [{\citenamefont {Campbell}\ and\ \citenamefont
  {Peden}(2005)}]{campbell_science_2005}%
  \BibitemOpen
  \bibfield  {author} {\bibinfo {author} {\bibfnamefont {C.~T.}\ \bibnamefont
  {Campbell}}\ and\ \bibinfo {author} {\bibfnamefont {C.~H.~F.}\ \bibnamefont
  {Peden}},\ }\href {\doibase 10.1126/science.1113955} {\bibfield  {journal}
  {\bibinfo  {journal} {Science}\ }\textbf {\bibinfo {volume} {309}},\ \bibinfo
  {pages} {713} (\bibinfo {year} {2005})},\ \Eprint
  {http://arxiv.org/abs/http://www.sciencemag.org/content/309/5735/713.full.pdf}
  {http://www.sciencemag.org/content/309/5735/713.full.pdf} \BibitemShut
  {NoStop}%
\bibitem [{\citenamefont {Fronzi}\ \emph {et~al.}(2009)\citenamefont {Fronzi},
  \citenamefont {Soon}, \citenamefont {Delley}, \citenamefont {Traversa},\ and\
  \citenamefont {Stampfl}}]{fronzi_jchemphys_2009}%
  \BibitemOpen
  \bibfield  {author} {\bibinfo {author} {\bibfnamefont {M.}~\bibnamefont
  {Fronzi}}, \bibinfo {author} {\bibfnamefont {A.}~\bibnamefont {Soon}},
  \bibinfo {author} {\bibfnamefont {B.}~\bibnamefont {Delley}}, \bibinfo
  {author} {\bibfnamefont {E.}~\bibnamefont {Traversa}}, \ and\ \bibinfo
  {author} {\bibfnamefont {C.}~\bibnamefont {Stampfl}},\ }\href {\doibase
  DOI:10.1063/1.3191784} {\bibfield  {journal} {\bibinfo  {journal} {Journal of
  Chemical Physics}\ }\textbf {\bibinfo {volume} {131}},\ \bibinfo {pages}
  {104701} (\bibinfo {year} {2009})}\BibitemShut {NoStop}%
\bibitem [{\citenamefont {Farmer}\ and\ \citenamefont
  {Campbell}(2010)}]{farmer_science_2010}%
  \BibitemOpen
  \bibfield  {author} {\bibinfo {author} {\bibfnamefont {J.~A.}\ \bibnamefont
  {Farmer}}\ and\ \bibinfo {author} {\bibfnamefont {C.~T.}\ \bibnamefont
  {Campbell}},\ }\href {\doibase 10.1126/science.1191778} {\bibfield  {journal}
  {\bibinfo  {journal} {Science}\ }\textbf {\bibinfo {volume} {329}},\ \bibinfo
  {pages} {933} (\bibinfo {year} {2010})},\ \Eprint
  {http://arxiv.org/abs/http://www.sciencemag.org/content/329/5994/933.full.pdf}
  {http://www.sciencemag.org/content/329/5994/933.full.pdf} \BibitemShut
  {NoStop}%
\bibitem [{\citenamefont {Nolan}(2010)}]{nolan_chemphyslett_2010}%
  \BibitemOpen
  \bibfield  {author} {\bibinfo {author} {\bibfnamefont {M.}~\bibnamefont
  {Nolan}},\ }\href {\doibase 10.1016/j.cplett.2010.09.016} {\bibfield
  {journal} {\bibinfo  {journal} {Chemical Physics Letters}\ }\textbf {\bibinfo
  {volume} {499}},\ \bibinfo {pages} {126} (\bibinfo {year}
  {2010})}\BibitemShut {NoStop}%
\bibitem [{\citenamefont {Galea}\ \emph {et~al.}(2009)\citenamefont {Galea},
  \citenamefont {Scanlon}, \citenamefont {Morgan},\ and\ \citenamefont
  {Watson}}]{galea_molsimulat_2009}%
  \BibitemOpen
  \bibfield  {author} {\bibinfo {author} {\bibfnamefont {N.~M.}\ \bibnamefont
  {Galea}}, \bibinfo {author} {\bibfnamefont {D.~O.}\ \bibnamefont {Scanlon}},
  \bibinfo {author} {\bibfnamefont {B.~J.}\ \bibnamefont {Morgan}}, \ and\
  \bibinfo {author} {\bibfnamefont {G.~W.}\ \bibnamefont {Watson}},\ }\href
  {\doibase 10.1080/08927020802707001} {\bibfield  {journal} {\bibinfo
  {journal} {Molecular Simulation}\ }\textbf {\bibinfo {volume} {35}},\
  \bibinfo {pages} {577} (\bibinfo {year} {2009})},\ \Eprint
  {http://arxiv.org/abs/http://www.tandfonline.com/doi/pdf/10.1080/08927020802707001}
  {http://www.tandfonline.com/doi/pdf/10.1080/08927020802707001} \BibitemShut
  {NoStop}%
\bibitem [{\citenamefont {Henderson}\ \emph {et~al.}(2003)\citenamefont
  {Henderson}, \citenamefont {Perkins}, \citenamefont {Engelhard},
  \citenamefont {Thevuthasan},\ and\ \citenamefont
  {Peden}}]{henderson_surfsci_2003}%
  \BibitemOpen
  \bibfield  {author} {\bibinfo {author} {\bibfnamefont {M.}~\bibnamefont
  {Henderson}}, \bibinfo {author} {\bibfnamefont {C.}~\bibnamefont {Perkins}},
  \bibinfo {author} {\bibfnamefont {M.}~\bibnamefont {Engelhard}}, \bibinfo
  {author} {\bibfnamefont {S.}~\bibnamefont {Thevuthasan}}, \ and\ \bibinfo
  {author} {\bibfnamefont {C.}~\bibnamefont {Peden}},\ }\href {\doibase
  10.1016/S0039-6028(02)02657-2} {\bibfield  {journal} {\bibinfo  {journal}
  {Surface Science}\ }\textbf {\bibinfo {volume} {526}},\ \bibinfo {pages} {1}
  (\bibinfo {year} {2003})}\BibitemShut {NoStop}%
\bibitem [{\citenamefont {Becke}\ and\ \citenamefont
  {Edgecombe}(1990)}]{becke_jchemphys_1990}%
  \BibitemOpen
  \bibfield  {author} {\bibinfo {author} {\bibfnamefont {A.~D.}\ \bibnamefont
  {Becke}}\ and\ \bibinfo {author} {\bibfnamefont {K.~E.}\ \bibnamefont
  {Edgecombe}},\ }\href@noop {} {\bibfield  {journal} {\bibinfo  {journal}
  {Journal of Chemical Physics}\ }\textbf {\bibinfo {volume} {92}},\ \bibinfo
  {pages} {5397} (\bibinfo {year} {1990})}\BibitemShut {NoStop}%
\end{thebibliography}%

\end{document}